\documentclass[prd,nofootinbib]{revtex4}
\everymath{\displaystyle}
\usepackage{amsfonts,amsmath,amssymb}
\usepackage{graphicx}

\newcommand{\SU}{\mathrm{SU}}
\newcommand{\SUc}{\mathrm{SU_c}}
\newcommand{\Nc}{N_\mathrm{c}}
\newcommand{\Nf}{N_\mathrm{f}}
\newcommand{\Nt}{N_\tau}
\newcommand{\Tc}{T_\mathrm{c}}
\newcommand{\Ttri}{T_\mathrm{tri}}
\newcommand{\Tend}{T_\mathrm{end}}
\newcommand{\muB}{\mu_\mathrm{B}}
\newcommand{\muI}{\mu_\mathrm{I}}
\newcommand{\mutri}{\mu_\mathrm{tri}}
\newcommand{\muend}{\mu_\mathrm{end}}
\newcommand{\rhoB}{\rho_\mathrm{B}}
\newcommand{\rhoI}{\rho_\mathrm{I}}
\newcommand{\arcsinh}{\mathrm{arcsinh}}
\newcommand{\arccosh}{\mathrm{arccosh}}

\def\Jvol<#1,#2,#3>{#1}
\def\Jpage<#1,#2,#3>{#2}
\def\Jyear<#1,#2,#3>{#3}
\newcommand\journal[1]{\textbf{\Jvol<#1>}, \Jpage<#1> (\Jyear<#1>)}
\newcommand\PRL[1]{Phys.\ Rev.\ Lett.\ \journal{#1}}

\newcommand\PRD[1]{Phys.\ Rev.\ D \journal{#1}}
\newcommand\PLB[1]{Phys.\ Lett.\ B \journal{#1}}
\newcommand\NPA[1]
{Nucl.\ Phys.\ \textbf{A\Jvol<#1>}, \Jpage<#1> (\Jyear<#1>)}
\newcommand\NPB[1]
{Nucl.\ Phys.\ \textbf{B\Jvol<#1>}, \Jpage<#1> (\Jyear<#1>)}
\newcommand\PR[1]{Phys.\ Rep.\ \journal{#1}}
\newcommand\PTP[1]{Prog.\ Theor.\ Phys.\ \journal{#1}}
\newcommand\PTPS[1]{Prog.\ Theor.\ Phys.\ Suppl.\ \journal{#1}}

\newcommand\JHEP[1]{J.\ High Energy Phys.\ \journal{#1}}
\newcommand\NC[1]{Nuovo Cimento \journal{#1}}
\newcommand\LNP[1]{Lect.\ Notes Phys.\ \journal{#1}}
\newcommand\PAN[1]{Phys.\ At.\ Nucl.\ \journal{#1}}
\newcommand\ibid[1]{\textit{ibid.}\ \journal{#1}}

\begin{document}

\title{Phase structures of strong coupling lattice QCD\\
       with finite baryon and isospin density}
\author{Yusuke~Nishida}
\affiliation{Department of Physics, University of Tokyo,
	     Tokyo 113-0033, Japan}
\date{\today}
\begin{abstract}
 Quantum chromodynamics (QCD) at finite temperature $(T)$, baryon
 chemical potential $(\muB)$ and isospin chemical potential $(\muI)$ is 
 studied in the strong coupling limit on a lattice with staggered
 fermions. With the use of large dimensional expansion and the mean 
 field approximation, we derive an effective action written in terms of 
 the chiral condensate and pion condensate as a function of $T$,
 $\muB$ and $\muI$. The phase structure in the space of $T$ and $\muB$ 
 is elucidated, and simple analytical formulas for the critical line of
 the chiral phase transition and the tricritical point are derived. 
 The effects of a finite quark mass $(m)$ and finite $\muI$
 on the phase diagram are discussed. We also investigate the
 phase structure in the space of $T$, $\muI$ and $m$, and 
 clarify the correspondence between color SU(3) QCD with finite
 isospin density and color SU(2) QCD with finite baryon
 density. Comparisons of our results with those from recent Monte Carlo
 lattice simulations on finite density QCD are given. 
\end{abstract}
\maketitle

\section{Introduction}

To reveal the nature of matter under extreme conditions in quantum
chromodynamics (QCD) is one of the most interesting and challenging
problems in hadron physics. 
At present the Relativistic Heavy Ion Collider (RHIC) experiments at
Brookhaven National Laboratory are running in order to create a new
state of hot matter, the quark-gluon plasma \cite{rhic,gyulassy03}. 
On the other hand, various novel states of dense matter have been
proposed at low temperature which are relevant to the physics of the 
interior of neutron star , such as the $^3\mathrm{P}_2$ neutron
superfluidity \cite{tamagaki70}, the pion \cite{migdal78} and kaon
\cite{kaplan86} condensations and so on. 
Furthermore, if the baryon density reaches as much as ten times the 
nuclear density at the core of the neutron stars, quark matter in a
color-superconducting state may be realized \cite{bailin84}. 

One of the first principles methods to solve QCD is numerical
simulation on the basis of the  Monte Carlo method. This has been
successfully applied to study the chiral and deconfinement phase
transitions at finite temperature ($T\neq0$) with zero baryon chemical
potential ($\muB=0$) \cite{karsch02}. 
On the other hand, lattice QCD simulations have an intrinsic 
difficulty at finite $\muB$ due to the complex fermion determinant. 
Recently, new approaches have been proposed to attack the problem in
$\SUc(3)$ QCD at finite $\muB$ \cite{muroya03}, such as the improved
reweighting method \cite{fodor02}, the Taylor expansion method around
$\muB =0$ \cite{allton02} and analytic continuation from imaginary
$\muB$ \cite{forcrand02,d'elia03}. They have indeed given us
some idea about the phase structure of QCD, e.g., the existence of
the critical end point of the chiral phase transition and the determination
of the slope of the critical line. They are, however, inevitably
restricted either to simulations on a small lattice volume or to the
low $\muB$ region near the critical line. 
Alternative attempts have also been made to study theories with positive
fermion determinant, which include color $\SUc(2)$ QCD with finite
$\muB$ \cite{hands99,kogut01} and color $\SUc(3)$ QCD with finite
isospin chemical potential ($\muI$) \cite{son01,kogut02}. 

In this paper we consider the strong coupling limit of  $\SUc(3)$   
lattice QCD
\cite{damgaard84,bilic92,bilic95,aloisio00,azcoiti03,fukushima03} 
and make  an extensive study of its phase structure with the use of the  
effective potential analytically calculated at finite $T$, $\muB$,
$\muI$ and the quark mass $m$. 
The phases obtained as functions of these variables are not easily
accessible in Monte Carlo simulations and thus give us some insights
into hot and dense QCD. Indeed, our previous study of strong coupling
$\SUc(2)$ QCD with finite $T$, $\muB$ and $m$ show results which agree
qualitatively well with those in numerical simulations \cite{nishida03}. 
Note that our approach has a closer connection to QCD than other
complementary approaches based on the Nambu$-$Jona-Lasinio 
model \cite{asakawa89}, the random matrix model \cite{halasz98} and 
the Schwinger-Dyson method \cite{takagi02,barducci90}.

This paper is organized as follows. In Sec.~\ref{sec:baryon},
we consider color $\SU(\Nc)$ QCD with finite temperature and
baryon density. Starting from the lattice action in the strong coupling
limit with one species of staggered fermion (corresponding to
$\Nf=4$ flavors in the continuum limit), we derive an effective free
energy in terms of a scalar mode $\sigma$ and a pseudoscalar mode 
$\pi$ for $\Nc\geq3$ using a large dimensional expansion and the mean
field approximation (Sec.~\ref{sec:baryon_formulation}). 
The resultant free energy is simple enough that one
can make analytical studies at least in the chiral limit $m=0$ with
finite $T$ and $\muB$. Secion \ref{sec:baryon_analytical} is devoted to
such studies and useful formulas for the critical temperature and
critical chemical potential for the chiral restoration are
derived. Inpaticular, we can present an analytical expression for the
temperature and baryon chemical potential of the tricritical point. 
In Sec.~\ref{sec:baryon_numerical}, we analyze the chiral condensate and
baryon density as functions of $\muB$ and show the phase diagram in the
plane of $T$ and $\muB$ for $\Nc=3, \Nf=4$. We discuss the effect of finite
quark mass $m$ on the phase diagram of QCD and comparison with the results
from recent lattice simulations are given here. 

In Sec.~\ref{sec:isospin}, we consider the strong coupling lattice QCD
with isospin chemical potential as well as $T$ and $\muB$. In
order to include $\muI$, we extend the formulation studied in
Sec.~\ref{sec:baryon} to two species of staggered fermion, which
corresponds to $\Nf=8$ flavor QCD in the continuum limit
(Sec.~\ref{sec:isospin_formulation}). Then we restrict ourselves to two
particular cases: One is at finite $T$, finite $\muB$ and \textit{small}
$\muI$ (Sec.~\ref{sec:isospin_case1}); the other is at finite $T$,
finite $\muI$ and \textit{zero} $\muB$ (Sec.~\ref{sec:isospin_case2}). 
For each case, we derive and analyze an analytical expression for the 
effective free energy and show the phase diagram in terms of $T$ and
$\muB$ or $\muI$ for $\Nc=3, \Nf=8$. In Sec.~\ref{sec:isospin_case1}, we 
discuss the effect of increasing $\Nf$ and that of
finite $\muI$ on the phase diagram of QCD in the $T$-$\muB$ plane. In 
Sec.~\ref{sec:isospin_case2}, comparison with the results from lattice 
simulations and discussion of a correspondence between QCD with finite
$\muI$ and $\SUc(2)$ QCD with finite $\muB$ are given. In Appendixes 
\ref{app:summation} and \ref{app:integration}, we give some technical
details for deriving the effective free energy.

\section{QCD ($\Nc\geq3,\Nf=4$) with finite baryon density 
 \label{sec:baryon}}

In this section, we consider strong coupling lattice QCD with finite 
baryon density. First of all, we review how to derive the effective
free energy written in terms of a scalar mode $\sigma$ and a
pseudoscalar mode $\pi$ for $\Nc\geq3$ for further extension in
Sec.~\ref{sec:isospin}, while the resultant free energy is almost the
same as that studied in \cite{damgaard84,bilic92,bilic95}. 
Then we analyze the chiral phase transition in the chiral limit at
finite temperature and density, and derive analytical formulas for
the second order critical line and the tricritical point. 
Finally we show the phase diagram in terms of $T$ and $\muB$ for $\Nc=3$.

\subsection{Formulation of strong coupling lattice QCD 
  \label{sec:baryon_formulation}}

In order to derive the effective free energy written in terms of the
scalar and pseudoscalar modes, we start from the lattice action with
one species of staggered fermion, which corresponds to the continuum QCD
with four flavors. In the strong coupling limit, the gluonic part of the
action vanishes because it is inversely proportional to the square of
the gauge coupling constant $g$. Consequently, the lattice action in the 
strong coupling limit is given by only the fermionic part:
\begin{align}
\begin{split}
 S[U,\chi,\bar\chi]=m\sum_x\bar\chi(x)\chi(x)
 &+\frac{1}{2}\sum_x\eta_0(x)\left\{\bar\chi(x)\mathrm{e}^\mu
 U_0(x)\chi(x+\hat{0})-\bar\chi(x+\hat{0})\mathrm{e}^{-\mu}
 U_0^\dagger(x)\chi(x)\right\}\\
 &+\frac{1}{2}\sum_x\sum_{j=1}^d\eta_j(x)
 \left\{\bar\chi(x)U_j(x)\chi(x+\hat{j})
 -\bar\chi(x+\hat{j})U_j^\dagger(x)\chi(x)\right\}\,.
\end{split}
 \label{eq:lattice_action}
\end{align}
$\chi$ stands for the quark field in the fundamental representation of
the color $\SU(\Nc)$ group and $U_\mu$ is the $\SU(\Nc)$
valued gauge link variable. $d$ represents the number of spatial
directions and we use a notation $x=(\tau,\vec{x})$ in which $\tau$
($\vec{x}$) represents the temporal (spatial) coordinate. $\eta_0(x)$
and $\eta_j(x)$ inherent in the staggered formalism are defined as
$\eta_0(x)=1$, $\eta_j(x)=(-1)^{\sum_{i=1}^j x_{i-1}}$.
$\mu$ is the quark chemical potential, while the temperature is defined
by $T=(aN_\tau)^{-1}$ with $a$ being the lattice 
spacing and $N_\tau$ being the number of temporal sites. We will write
all the dimensionful quantities in units of $a$ and will not write $a$
explicitly.

This lattice action has
$\mathrm{U}(1)_\mathrm{V}\times\mathrm{U}(1)_\mathrm{A}$ symmetry
in the chiral limit $m=0$, which is a remnant of the four flavor chiral 
symmetry in the continuum theory, defined by 
$\mathrm{U}(1)_\mathrm{V}:
\chi(x)\mapsto\mathrm e^{\mathrm i\theta_\mathrm{V}}\chi(x),\ 
\bar\chi(x)\mapsto\bar\chi(x)\mathrm e^{-\mathrm i\theta_\mathrm{V}}$
and 
$\mathrm{U}(1)_\mathrm{A}:
\chi(x)\mapsto\mathrm
e^{\mathrm i\varepsilon(x)\theta_\mathrm{A}}\chi(x),\ 
\bar\chi(x)\mapsto\bar\chi(x)\mathrm
e^{\mathrm i\varepsilon(x)\theta_\mathrm{A}}$.
Here $\varepsilon(x)$ is given by
$\varepsilon(x)=(-1)^{\sum_{\nu=0}^{d}x_{\nu}}$, 
which plays a similar role to $\gamma_5$ in the continuum
theory. $\mathrm{U}(1)_\mathrm{A}$ will be explicitly broken by
the introduction of a finite quark mass or spontaneously broken by
condensation of $\left\langle\bar\chi\chi\right\rangle$. Note that the
staggered fermion's $\mathrm{U}(1)_\mathrm{A}$ symmetry should not be
confused with the axial $\mathrm{U}(1)$ symmetry in the continuum
theory, which is broken by the quantum effect. 
Now by using the lattice action Eq.~(\ref{eq:lattice_action}), we can
write the partition function as 
\begin{align}
Z=\int\mathcal D[\chi,\bar\chi]\,\mathcal D[U_0]\,\mathcal D[U_j]\,
\mathrm e^{-S}\,.
\end{align}
In the succeeding part of this subsection, we perform the path integrals by
use of the large dimensional $(1/d)$ expansion and the mean field
approximation to derive the effective free energy. 

First, we perform the integration over spatial gauge link variable $U_j$ 
using the formulas for the $\mathrm{SU}(\Nc)$ group integration. 
Keeping only the lowest order term of the Taylor expansion of
$\mathrm{e}^{-S}$, which corresponds to the leading order term of the
$1/d$ expansion \cite{kluberg-stern83}, 
we can calculate the group integral as follows:
\begin{align}
 \int\mathcal D[U_j]\,\exp\left[-\frac12\sum_x\sum_{j=1}^d\eta_j(x)
 \left\{\bar\chi(x)U_j(x)\chi(x+\hat j)
 -\bar\chi(x+\hat j)U_j^\dagger(x)\chi(x)\right\}\right]
 =\exp\left[\sum_{x,y}M(x)V_\mathrm{M}(x,y)M(y)\right]+\cdots\,,
 \label{eq:1/d_expansion}
\end{align}
where $M(x)$ and $V_\mathrm{M}(x,y)$ are the mesonic composite and
its propagator in the spatial dimensions, defined, respectively, by
\begin{align}
 M(x)=\frac1\Nc\delta_{ab}\,\bar\chi^a(x)\chi^b(x)\,,\qquad
 V_\mathrm{M}(x,y)=\frac\Nc8\sum_{j=1}^d
 \left(\delta_{y,x+\hat j}+\delta_{y,x-\hat j}\right)\,. 
\end{align}
The resultant term in the action describes the nearest neighbor
interaction of the mesonic composite $M(x)$. Residual terms are in 
higher order of the $1/d$ expansion and vanish in the large dimensional 
limit. Note that Eq.~(\ref{eq:1/d_expansion}) is correct as long as
$\Nc\geq3$. For $\Nc=2$ the baryonic composite (diquark
field) gives the same contribution as the mesonic one in the $1/d$ 
expansion. The finite density $\SUc(2)$ QCD with a
diquark field was studied in a previous paper \cite{nishida03}. 

Next, we linearize the four-fermion term in
Eq.~(\ref{eq:1/d_expansion}) by introduction of the auxiliary field
$\sigma(x)$. This is accomplished by the standard Gaussian technique; 
\begin{align}
 \exp\left[\sum_{x,y}M(x)V_\mathrm{M}(x,y)M(y)\right]
 =\int\mathcal D[\sigma]\,\exp\left[-\sum_{x,y}\left\{
 \sigma(x)V_\mathrm{M}(x,y)\sigma(y)
 +2\sigma(x)V_\mathrm{M}(x,y)M(y)\right\}\right]\,.
 \label{eq:auxiliary_sigma}
\end{align}
From the above transformation, we can show the relation between the 
vacuum expectation values of the auxiliary field and the mesonic
composite: 
$\left\langle\sigma(x)\right\rangle=-\left\langle M(x)\right\rangle$. 
In order to keep the action invariant under the
$\mathrm{U}(1)_\mathrm{V}$ and $\mathrm{U}(1)_\mathrm{A}$
transformations, we should assign the transformation for the
auxiliary field as $\mathrm{U}(1)_\mathrm{V}:\sigma(x)\mapsto\sigma(x)$ 
and $\mathrm{U}(1)_\mathrm{A}:\sigma(x)\mapsto
\mathrm e^{2\mathrm{i}\varepsilon(x)\theta_\mathrm{A}}\sigma(x)$.

In the staggered fermion formalism, one species of staggered fermion at 
various sites is responsible for the degree of freedom of flavors and
spins. Therefore the auxiliary field $\sigma(x)$ is responsible for both 
the scalar mode $\sigma$ and the pseudoscalar mode $\pi$ 
\cite{kluberg-stern83}. 
In order to derive the effective free energy written using the scalar and
pseudoscalar modes, we introduce $\sigma$ and
$\pi$ as $\sigma(x)=\sigma+\mathrm{i}\varepsilon(x)\pi$ 
\cite{kawamoto81}. 
Then we can find a correspondence between these new values $\sigma$,
$\pi$ and the vacuum expectation values of the mesonic composites:
$\sigma=-\left\langle\bar\chi^a(x)\chi^a(x)\right\rangle/\Nc$, 
$\pi=\left\langle\mathrm{i}\,\bar\chi^a(x)\varepsilon(x)
\chi^a(x)\right\rangle/\Nc$.
These equations tell us that $\sigma$ and $\pi$ are
even and odd under lattice parity, respectively \cite{hands99,doel83}, 
and therefore $\sigma$ and $\pi$ correspond to the spatially uniform
condensates of the scalar mode (chiral condensate) and the pseudoscalar
mode (pion condensate). They transform under $\mathrm{U}(1)_\mathrm{A}$
transformation as 
\begin{align}
 \mathrm{U}(1)_\mathrm{A}:\quad
 \begin{pmatrix}
  \sigma\\\pi
 \end{pmatrix}
 \mapsto
 \begin{pmatrix}
  \cos2\theta_\mathrm{A}&-\sin2\theta_\mathrm{A}\\
  \sin2\theta_\mathrm{A}&\phantom{-}\cos2\theta_\mathrm{A}
 \end{pmatrix}
 \begin{pmatrix}
  \sigma\\\pi
 \end{pmatrix}\,.
\end{align}
The partition function is then written as
\begin{align}
  Z&=\int\mathcal{D}[U_0]\,\mathcal{D}[\chi,\bar\chi]\,
  \mathrm{e}^{-S'[U_0,\chi,\bar\chi;\sigma,\pi]}\,,
\label{eq:partition}
\end{align}
with
\begin{align}
 \begin{split}
  S'[U_0,\chi,\bar\chi;\sigma,\pi]
  =&\sum_x\left[m\bar\chi(x)\chi(x)+\frac{1}{2}\left\{\bar\chi(x)
  \mathrm{e}^\mu U_0(x)\chi(x+\hat0)-\bar\chi(x+\hat0)
  \mathrm{e}^{-\mu}U_0^\dagger(x)\chi(x)\right\}\right.\\
  &\quad\qquad\qquad\qquad
  \left.+\frac{\Nc d}{4}\left(\sigma^2+\pi^2\right)+\frac{\Nc d}2
  \left\{\sigma-\mathrm{i}\varepsilon(x)\pi\right\}M(x)\right]\,. 
 \end{split}
\label{eq:prime_action}
\end{align}
Note that this action contains hopping terms of $\chi$, $\bar\chi$ 
only in the temporal direction after the $1/d$ expansion and
the mean field approximation.

In order to complete the remaining integrals, we adopt the Polyakov
gauge \cite{damgaard84} in which $U_0(\tau,\vec{x})$ is diagonal and
independent of $\tau$:
\begin{align}
 U_0(\tau,\vec{x})=\mathrm{diag}
 \left[\mathrm{e}^{\mathrm{i}\phi_1(\vec{x})/N_\tau},\dots,
 \mathrm{e}^{\mathrm{i}\phi_{\Nc}(\vec{x})/N_\tau}\right]\quad
 \qquad\text{with}\quad\sum_{a=1}^{\Nc}\phi_a(\vec{x})=0\,.
\label{eq:gauge_choice}
\end{align}
Also we make a partial Fourier transformation for the quark fields;
\begin{align}
 \chi(\tau,\vec{x})=\frac{1}{\sqrt{N_\tau}}\sum_{n=1}^{N_\tau}
 \mathrm{e}^{\mathrm{i}k_n\tau}\tilde\chi(n,\vec{x})\,,\quad
 \bar\chi(\tau,\vec{x})=\frac{1}{\sqrt{N_\tau}}\sum_{n=1}^{N_\tau}
 \mathrm{e}^{-\mathrm{i}k_n\tau}\tilde{\bar\chi}(n,\vec{x})\,,
 \qquad k_n=2\pi\frac{(n-\frac{1}{2})}{N_\tau}\,.
\label{eq:fourier_trans}
\end{align}
Substituting Eqs.~(\ref{eq:gauge_choice}) and (\ref{eq:fourier_trans})
into the action (\ref{eq:prime_action}) and taking the summation over
$\tau$, the Grassmann integration over the the quark fields $\tilde\chi$ 
and $\tilde{\bar\chi}$ results in the following determinant:
\begin{align}
 \prod_{\vec{x}}\prod_{a=1}^{\Nc}\prod_{n=1}^{N_\tau/2}\left[
 \sin^2\bar k_n+M^2+\left(\frac d2\pi\right)^2\right]
\end{align}
with $\bar k_{n}=k_n+\phi_a(\vec{x})/N_\tau-\mathrm{i}\mu$.
$M$ denotes the dynamical quark mass defined by
$M=m+(d/2)\sigma$. 

The product with respect to the Matsubara frequencies $n$ can be
performed using a technique similar to that in the calculation of the
free energy in finite temperature continuum field theory. The
details of the calculation are given in Appendix~\ref{app:summation}. 
The result turns out to have a quite simple form:
$2\cosh\left[N_\tau E\right]
+2\cos\left[\phi_a(\vec{x})-\mathrm{i}N_\tau\mu\right]$ 
with one-dimensional quark excitation energy
$\textstyle E[\sigma,\pi]=\arcsinh[\sqrt{M^2+(d/2)^2\pi^2}]$.

Finally we can perform the integration with respect to $U_0$ 
by use of the formula in Appendix~\ref{app:integration}
for the Polyakov gauge. Then the integration gives, up to an irrelevant 
factor, 
\begin{align}
 \int\mathcal{D}[U_0]\prod_{\vec{x}}\prod_{a=1}^{\Nc}
 \left\{2\cosh\left[N_\tau E\right]
 +2\cos[\phi_a(\vec{x})-\mathrm{i}N_\tau\mu]\right\}
 =\prod_{\vec{x}}\left\{\sum_{n}\det_{i,j}P_{n+i-j}\right\}\,,
\end{align}
where $P_0=2\cosh\left[N_\tau E\right]$, 
$P_{\pm1}=\cosh\left[N_\tau\mu\right]\pm\sinh\left[N_\tau\mu\right]
=\mathrm{e}^{\pm\Nt\mu}$ and
$P_{|n|\geq3}=0$. The determinant is to be taken with respect to
$i,j=1,2,\dots,\Nc$. The determinants have nonvanishing
values only for $n=0, \pm1$ and their summation can be calculated
exactly for general $\Nc$. First, for $n=0$, $\det P_{i-j}$ is
expressed in the form of an $\Nc\times\Nc$ matrix as
\begin{align}
 \det_{i,j}P_{i-j}=
 \begin{vmatrix}
  P_0&P_{-1}&0&\cdots\\
  P_{+1}&P_0&\ddots&\\
  0&\ddots&\ddots&P_{-1}\\
  \vdots&&P_{+1}&P_0
 \end{vmatrix}\,.
\end{align}
Making a recursion formula for it and using the fact that
$P_{-1}P_{+1}=1$, we can obtain the simple solution 
$\det P_{i-j}=\sinh\left[(\Nc+1)N_\tau E\right] 
/\sinh\left[N_\tau E\right]$ \cite{damgaard84}. 
Then the calculation for $n=\pm1$ is
rather easy and results in $\det P_{-1+i-j}+\det P_{1+i-j}
=P_{-1}^{\,\Nc}+P_{+1}^{\,\Nc}=2\cosh\left[\Nc N_\tau\mu\right]$.
As a result, the effective free energy is given as follows;
\begin{align}
 F_\mathrm{eff}[\sigma,\pi;T,\muB]
 =\left(-\log Z\right)/\left({\textstyle\sum_x}\right)
 =\frac{\Nc d}{4}\left(\sigma^2+\pi^2\right)
 -T\log\left\{2\cosh\left[\muB/T\right]
 +\frac{\sinh\left[(\Nc+1)E/T\right]}{\sinh\left[E/T\right]}\right\}\,.
 \label{eq:free_energy}
\end{align}
Here we have defined the baryon chemical potential $\muB$ as 
$\muB=\Nc\mu$. Although the temperature $T=1/\Nt$ takes discrete values,
the effective free energy $F_\mathrm{eff}[1/\Nt]$ can be uniquely
extended to a function of continuous $T$ in terms of an analytic
continuation since $F_\mathrm{eff}[1/\Nt]$ is defined on an infinite
sequence of points. Hereafter we consider $T$ as a continuous variable.
This effective free energy in Eq.~(\ref{eq:free_energy}) corresponds to the
$\Nf=4$ flavor QCD in the continuum limit.

\subsection{Analytical properties in the chiral limit
  \label{sec:baryon_analytical}}

The effective free energy obtained in the previous subsection
is simple enough to make analytical studies
in the chiral limit. This is one of the main advantages of our
approach. Such analytical studies on the chiral phase transition
are useful in understanding the phase structure of strong coupling
lattice QCD, which will be presented in the next subsection. 
In the chiral limit $m=0$, the quark excitation energy reduces to
$\textstyle E[\sigma,\pi]=
\arcsinh\left[\left(d/2\right)\sqrt{\sigma^2+\pi^2}\right]$.
Therefore the effective free energy in Eq.~(\ref{eq:free_energy}) is
a function of only $\textstyle\sqrt{\sigma^2+\pi^2}$. Of course
this symmetry comes from the chiral symmetry $\mathrm{U}(1)_\mathrm{A}$
of the original lattice action. Since we can arbitrarily choose the
direction of the condensate, we take $\sigma\neq0$ and $\pi=0$.

\subsubsection*{Chiral restoration at finite temperature}

As we will confirm numerically in the next subsection, the effective
free energy exhibits a second order phase transition at finite
temperature. Therefore we can expand it in terms of the order parameter
$\sigma$ near the critical point. Expansion of the effective free energy
up to second order of the chiral condensate $\sigma$ gives
\begin{align}
 &F_{\mathrm{eff}}[\sigma,0]
 =\frac{\Nc d}{4}\sigma^2-\frac{d^2\Nc(\Nc+1)(\Nc+2)}
 {24T(\Nc+1+2\cosh\left[\muB/T\right])}\sigma^2
 +O\left(\sigma^4\right)\,.
\end{align}
As long as the coefficient of $\sigma^4$ is positive, the condition that
the coefficient of $\sigma^2$ is zero gives the critical chemical
potential for the second order phase transition as a function of
temperature:
\begin{align}
 \muB^\mathrm{cri}(T)=T\,\arccosh
 \left[\frac{(\Nc+1)\{d(\Nc+2)-6T\}}{12T}\right]\,.
 \label{eq:critical_mu}
\end{align}
The critical temperature at zero chemical potential is given by solving
the equation $\muB^\mathrm{cri}(T)=0$ and turns out to be
$T_\mathrm{c}(0)=d(\Nc+1)(\Nc+2)/[6(\Nc+3)]$.

When the coefficient of $\sigma^4$ becomes zero, the phase transition
becomes of first order. An analytical expression for the tricritical point  
$(T_\mathrm{tri},\mu_\mathrm{tri})$ can be obtained as a solution of
the coupled equations of Eq.~(\ref{eq:critical_mu}) and 
$\partial^4F_\mathrm{eff}/\partial\sigma^4=0$, which results in
\begin{align}
 T_\mathrm{tri}
 =\frac{\sqrt{225\Nc^2+20d^2(3\Nc^2+6\Nc-4)}-15\Nc}{20d}\,,\quad
 \mu_\mathrm{tri}=\muB^\mathrm{cri}(T_\mathrm{tri})\,.
 \label{eq:tricritical}
\end{align}
The effective free energy exhibits a second order chiral phase
transition for $T\geq T_\mathrm{tri}$ and it becomes of first order
for $T<T_\mathrm{tri}$. The existence of the tricritical point is
consistent with the results of other analytical approaches 
using the Nambu$-$Jona-Lasinio model \cite{asakawa89}, 
random matrix model \cite{halasz98}, Schwinger-Dyson equation 
\cite{takagi02} and other methods \cite{barducci90,antoniou03,hatta03}.

\subsubsection*{Chiral restoration at finite density}

An analytical study of the first order chiral phase transition at finite
temperature for $T<T_\mathrm{tri}$ is rather involved. However, the
effective free energy reduces to a simpler form at $T=0$:
\begin{align}
 F_\mathrm{eff}[T=0]=\frac{\Nc d}{4}\sigma^2
 -\mathrm{max}\left\{\muB,\Nc E[\sigma]\right\}\,,\quad
 E[\sigma]=\arcsinh\left[\frac d2\left|\sigma\right|\right]\,.
 \label{eq:T=0}
\end{align}
It is easy to study the first order phase transition analytically in
this case. 
 
The effective free energy has two local minima as a function of
$\sigma$: One is $F_\mathrm{eff}=-\muB$ at $\sigma=0$;
the other is $F_\mathrm{eff}=\Nc d\,\sigma_0^{\,2}/4-\Nc E[\sigma_0]<0$
at $\sigma=\sigma_0$, where $\sigma_0$ is the solution of the chiral gap 
equation $\partial F_\mathrm{eff}/\partial\sigma=0$,
\begin{align}
 \sigma_0^{\,2}\left[1+\left(\frac d2\sigma_0\right)^2\right]
 =1\,,\qquad\ \text{therefore}\quad
 \sigma_0^{\,2}=\frac{2\sqrt{1+d^2}-2}{d^2}\,. 
\label{eq:sigma0}
\end{align}
As $\muB$ becomes larger, the global minimum changes
from $\Nc d\,\sigma_0^{\,2}/4-\Nc E[\sigma_0]$ to $-\muB$ at some value 
of the chemical potential. This is the critical chemical potential,
given by $\muB^\mathrm{cri}(T=0)=\Nc E[\sigma_0]-\Nc
d\,\sigma_0^{\,2}/4$. 
At this critical point, the order parameter $\sigma$ changes
discontinuously as 
\begin{align}
 \sigma(\mu)=
 \begin{cases}
  \,\sigma_0\frac{}{}\quad&\text{when\ \ }\muB<\muB^\mathrm{cri}\,,\\
  \,\,0\frac{}{}&\text{when\ \ }\muB>\muB^\mathrm{cri}\,.
 \end{cases}
 \label{eq:chiral}
\end{align}
This is a first order phase transition. We can also easily calculate
the baryon density $\rhoB$ at $T=0$, and it has a discontinuity
associated with that of the chiral condensate:
\begin{align}
 \rhoB=-\frac{\partial F_\mathrm{eff}}{\partial\muB}=
 \begin{cases}
  \,\,0\frac{}{}&\text{when\ \ }\muB<\muB^\mathrm{cri}\,,\\
  \,\,1\quad&\text{when\ \ }\muB>\muB^\mathrm{cri}\,.
 \end{cases}
 \label{eq:density}
\end{align}
At the critical chemical potential, $\rhoB$ changes from the 
empty density $0$ to the saturated density $1$ \cite{aloisio00}. 
Note that one baryon
(= $\Nc$ quarks) at one lattice site, that is $\rhoB=1$, is the
maximally allowed configuration by the Fermi statistics for one species
of staggered fermion. Saturation at large chemical potential on the 
lattice has also been observed in lattice simulations and strong
coupling analyses of $\SUc(2)$ QCD \cite{kogut01,nishida03}.

This chiral phase transition with finite $\muB$ at
$T=0$ is not associated with the deconfinement phase transition to
quark matter. Equations (\ref{eq:chiral}) and
(\ref{eq:density}) show that the chiral restoration coincides
with the transition from vacuum to saturated nuclear matter, 
although they are in general separated \cite{halasz98}. 
Including the contribution of dynamical baryons, which is suppressed in
the leading order of the $1/d$ expansion, will introduce the Fermi
surface of baryons and hence separate the two phase transitions.

\subsection{The phase structure with finite quark mass 
  \label{sec:baryon_numerical}}

In this subsection, setting the number of colors and spatial dimensions
to 3, we examine numerically the nature of the chiral
phase transition and the phase diagram of strong coupling lattice QCD
for $\Nc=3,\Nf=4$.

\subsubsection*{Chiral condensate and baryon density}

\begin{figure}[tp]
 \begin{center}
  \includegraphics[width=0.95\textwidth,clip]{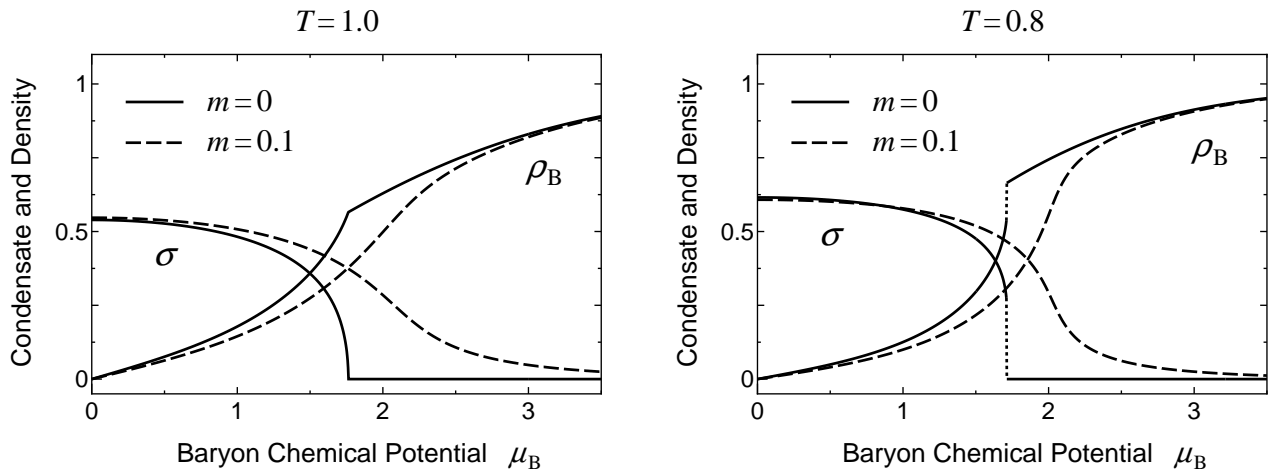}
  \caption{Chiral condensate $\sigma$ and baryon density $\rhoB$ as
  functions of the baryon chemical potential $\muB$ for
  $\Nc=3,\Nf=4$. In the left panel, they are plotted just above the
  tricritical temperature ($\Ttri=0.866$) at $T=1.0$ for $m=0$ (the
  solid line) and $m=0.1$ (the dashed one). In the right panel, they are
  plotted just below $\Ttri$ at $T=0.8$ for $m=0$ (the solid line) and
  $m=0.1$ (the dashed one). 
  \label{fig:cond_dens}}
 \end{center}
\end{figure}

By minimizing the effective free energy $F_\mathrm{eff}$ for $\Nc=3$,
$d=3$ in terms of the order parameter $\sigma$, we determine the chiral
condensate as a function of $T$ and $\muB$. The baryon density $\rhoB$
is also calculated by $-\partial F_\mathrm{eff}/\partial\muB$. 
In Fig.~\ref{fig:cond_dens}, the results with zero and finite $m$
are shown as functions of $\muB$. 

In the left panel of Fig.~\ref{fig:cond_dens}, $\sigma$ and $\muB$ are
shown for $T=1.0$ just above $\Ttri=0.866$. In the chiral limit they
show a second order phase transition at the critical chemical potential
given by Eq.~(\ref{eq:critical_mu}), while introduction of an
infinitesimal $m$ makes the transition a smooth crossover. 

$\sigma$ and $\muB$ are shown for $T=0.8$ just below $\Ttri$ in the
right panel of Fig.~\ref{fig:cond_dens}. As discussed in the previous
subsection, they show jumps at the critical chemical potential in the
chiral limit. This first order phase transition is weakened by the
introduction of $m$ and finally becomes a crossover for large quark
masses.

\subsubsection*{The phase diagram}

\begin{figure}[tp]
 \begin{center}
  \includegraphics[width=0.9\textwidth,clip]{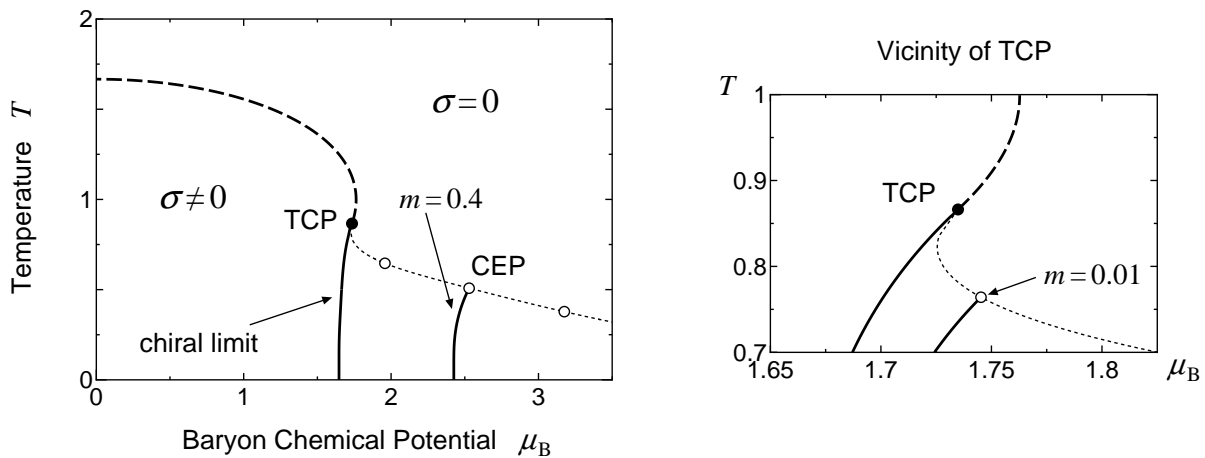}
  \caption{Phase diagram of strong coupling lattice QCD with
  $\Nc=3,\Nf=4$ in the $T$-$\muB$ plane. The solid (dashed) line denotes
  the phase transition line 
  for the first (second) order phase transition in the chiral limit. The 
  black circle represents the tricritical point (TCP) where the nature 
  of the phase transition changes from first order to 
  second. The first order line and its critical end point (CEP) for
  $m=0.4$ are also shown. The dotted line denotes the flow of the
  critical end points when the quark mass is enlarged from 0. White
  circles are put on the critical end points for $m=0.1$ (the left one),
  $m=0.4$ (the middle) and $m=0.8$ (the right one) in the left
  panel. The vicinity of the TCP is enlarged and the first order line and
  CEP for $m=0.01$ are shown in the right panel. 
  \label{fig:phase_diagram}}
  \end{center}
\end{figure}

Now we show in Fig.~\ref{fig:phase_diagram} the phase diagram of the
strong coupling lattice QCD with $\Nc=3,\Nf=4$ in the plane of
$T$ and $\muB$. In the chiral limit, the chiral condensate $\sigma$ has 
nonvanishing value at low $T$ and small $\muB$, 
which means a spontaneous breakdown of the chiral symmetry
$\mathrm{U_A}(1)$ by the condensation $\sigma\neq0$. The symmetry broken 
phase is separated by a phase transition from the chiral symmetric phase;
the phase transition is of second order in the high temperature region
while it is of first order in the low temperature region. 
The second order critical line merges smoothly into the first order one at
the tricritical point 
$(T_\mathrm{tri}, \mu_\mathrm{tri})=(0.866, 1.73)$.

In contradiction to our result, the chiral phase transition of
four-flavor QCD in the chiral limit is thought to be of first order
\cite{fodor02,d'elia03}. 
This difference comes from the fact that the symmetry of the
staggered fermion action is limited to
$\mathrm{U}(1)\times\mathrm{U}(1)$ in the strong coupling limit and we 
have employed the mean field approximation.\footnote{As one gets closer
to the continuum limit, the symmetry will turn into the usual
four-flavor chiral symmetry. Then the chiral fluctuations may induce a
first order phase transition, as they do in the corresponding linear
sigma model \cite{pisarski84}.}
However, it is sensible to consider that the second order critical line
in our result corresponds to the phase transition line of chiral
restoration in QCD. 

As we can see from Fig.~\ref{fig:phase_diagram}, 
the critical temperature falls much more rapidly as a function of 
$\muB$ in contrast to the results from recent Monte Carlo simulations
at small chemical potential
\cite{fodor02,allton02,forcrand02,d'elia03}. 
In order to make the discussion quantitative, we calculate the slope of
the critical temperature near zero chemical potential. 
From Eq.~(\ref{eq:critical_mu}), 
\begin{align}
 \frac{\Tc(\mu)}{T_\mathrm{c}(0)}
 =1-1.5\frac{\mu^2}{T_\mathrm{c}(0)^2}-3.375\frac{\mu^4}{T_\mathrm{c}(0)^4} 
 +O\left(\left(\mu/T_\mathrm{c}\right)^6\right)
 \label{eq:slope}
\end{align}
with $T_\mathrm{c}(0)=5/3$. Here we have written
Eq.~(\ref{eq:slope}) in terms of the quark chemical potential
$\mu=\muB/\Nc$ in order to make the comparison to the lattice data
easy. The value of the coefficient of the second 
term is two orders of magnitude larger than that calculated in the
lattice simulation with $\Nf=4$ for $m=0.05$ \cite{d'elia03}. 
This large difference in the slope of the chiral phase transition may be 
understood due to the effect of $m=0$ and that of $g\to\infty$. 
Our expression (\ref{eq:slope}) is derived in the chiral limit, 
while all of the lattice simulations are performed with finite quark 
mass. As reported in \cite{karsch03}, the chiral transition line
has significant quark mass dependence in the lattice simulations: The
slope of $\Tc(\mu)$ suffers a sharp increase on decreasing $m$, from
0.025 ($m=0.1$) to 0.114 ($m=0.005$) in $\Nf=3$ QCD. 
In addition to that, as we can find in \cite{bilic95},
the introduction of the next to leading order term of $1/g^2$ and $1/d$
in strong coupling lattice QCD decreases the critical temperature, and
the amount of decrease at zero chemical potential is more than that at
finite chemical potential. Therefore the slope of the critical
temperature will be made softer by the corrections for the strong
coupling and large dimensional expansions. Combining the above two
effects, the difference of the slope between our result and lattice
simulations could become reasonably small; this needs further study.

Next we consider the effect of quark mass on the phase diagram. 
Introduction of finite $m$ washes out the second order phase
transition as shown in the left panel of Fig.~\ref{fig:cond_dens}. As a 
result, the first order phase transition line terminates at a second
order critical end point $(\Tend, \muend)$. As long as the quark mass is
very small, $m\lesssim0.001$, the critical end point flows almost along
the tangent line at the tricritical 
point as in the right panel of Fig.~\ref{fig:phase_diagram}. 
Then for the larger quark mass $m\gtrsim0.001$, $\Tend$ decreases and
$\muend$ increases. Also, the first order phase transition line shifts in 
the direction of large $\muB$ associated with this change. 

Strong coupling lattice QCD may give us some indication for
lattice simulations about the effect of $m$ on the
phase diagram of QCD. As shown in Table~\ref{tab:critical_points}, 
a finite quark mass $m=0.1$ decreases $\Tend$ by 25\% and  
increases $\muend$ by 12\% from the tricritical point 
$(\Ttri, \mutri)$. We can find in the
same way that even a quark mass as small as $m=0.001$ can move the
critical end point by 5\%. This indicates the relatively large
dependence of the phase diagram of QCD on the quark mass.

Our phase structure is consistent with those from other analytical
approaches 
\cite{asakawa89,halasz98,takagi02,barducci90,antoniou03,hatta03}, 
except for one main difference \cite{fukushima03}. This is the fact
that the gradient of the phase transition line near the tricritical
point or critical end point is positive. Consequently, the critical end
point flows in the direction of smaller 
$T$ and smaller $\muB$ as a function of small quark mass.
The positive gradient of the chiral phase transition line near the
tricritical point can be understood as follows. The gradient
$\mathrm{d}T/\mathrm{d}\mu$ is related to  
discontinuities in baryon density $\Delta\rho$ and in entropy
density $\Delta s$ across the phase boundary through the
generalized Clapeyron-Clausius relation \cite{halasz98}
\begin{align}
 \frac{\mathrm{d}T}{\mathrm{d}\mu}=-\frac{\Delta\rho}{\Delta s}\,.
\end{align}
As we can see in the right panel of Fig.~\ref{fig:cond_dens}, 
the baryon density in the chiral restored phase $\rho_+$ is larger
than that in the chiral broken phase $\rho_-$ just below the
tricritical point $(T=0.8, \muB=1.7)$. We can also calculate the entropy
density from the effective free energy (\ref{eq:free_energy}) showing
the result that $\Delta s<0$. $\Delta\rho=\rho_+-\rho_->0$ holds because
baryons are massless in the chiral restored phase and we can fill
more baryons under fixed $\muB$. Since we are in the leading order of
the $1/d$ expansion, all baryons are static. Then we can estimate the
entropy as 
$s_\pm\sim-\rho_\pm\log\rho_\pm-(1-\rho_\pm)\log(1-\rho_\pm)$. Because 
$1>\rho_+>\rho_->0.5$, the discontinuities in the entropy density is
negative $\Delta s=s_+-s_-<0$. Therefore the gradient of
the chiral phase transition line just below the tricritical point turns
out to be positive, $\mathrm{d}T/\mathrm{d}\mu=-\Delta\rho/\Delta s>0$.

\begin{table}[tp]
 \begin{center}
  \caption{Tricritical point $(T_\mathrm{tri}, \mu_\mathrm{tri})$ for
  $m=0$ and critical end points $(T_\mathrm{end}, \mu_\mathrm{end})$
  for various small quark masses. The deviations of the
  critical end points from the tricritical point are also shown.
  \label{tab:critical_points}}\medskip
  \begin{tabular}{lccrr}\hline\hline
   \multicolumn{1}{c}{$\quad m\quad$}
   &$\quad T_\mathrm{end}\quad$&$\quad\mu_\mathrm{end}\quad$ 
   &\multicolumn{1}{c}{$(T_\mathrm{end}-T_\mathrm{tri})/T_\mathrm{tri}$}
   &\multicolumn{1}{c}{$(\mu_\mathrm{end}-\mu_\mathrm{tri})/\mu_\mathrm{tri}$}
   \\\hline
   \ 0&0.866&1.73&\multicolumn{1}{c}{---}&\multicolumn{1}{c}{---}\\
   \ 0.001&0.823&1.73&$-5.03\,\%$$\qquad$
   &$-0.541\,\%$$\qquad$\\
   \ 0.01&0.764&1.75&$-11.8\,\%$\phantom{0}$\qquad$
   &\phantom{0}0.599\,\%$\qquad$\\
   \ 0.05&0.690&1.85&$-20.4\,\%$\phantom{0}$\qquad$
   &\phantom{0}6.37\,\%\phantom{0}$\qquad$\\
   \ 0.1&0.646&1.96&$-25.5\,\%$\phantom{0}$\qquad$
   &12.9\,\%\phantom{00}$\qquad$\\
   \ 0.2&0.590&2.16&$-31.9\,\%$\phantom{0}$\qquad$
   &24.7\,\%\phantom{00}$\qquad$\\
  \hline\hline\end{tabular}
 \end{center}
\end{table}

\section{QCD ($\Nc\geq3,\Nf=8$) with finite isospin density
 \label{sec:isospin}}

In this section, we consider strong coupling lattice QCD with
finite isospin density. First of all, we derive the analytical
expression for the effective free energy in two particular cases. One is 
at finite $T$, finite $\muB$ and \textit{small} $\muI$ with chiral
condensates. The other is at finite $T$, finite $\muI$ and \textit{zero}
$\muB$ with chiral and pion condensates. Then we analyze the
effective free energy and show the phase diagrams in terms of
$T$ and $\muB$ or $\muI$ for $\Nc=3,\Nf=8$.

\subsection{Strong coupling lattice QCD with isospin chemical potential
  \label{sec:isospin_formulation}}

In order to investigate the effect of the isospin chemical
potential $\mu_\mathrm{I}$ on the phase structure of QCD, 
we have to extend the lattice action studied in the
previous section to that with two species of staggered fermion having
degenerate masses. The lattice action with two species of staggered
fermion corresponds to the eight-flavor QCD in the continuum limit. Then 
for $m=0$ and $\muI=0$, it has
$\mathrm{U}(2)_\mathrm{V}\times\mathrm{U}(2)_\mathrm{A}$ symmetry, 
as a remnant of eight-flavor chiral symmetry, defined by
$\chi(x)\mapsto\mathrm e^{\mathrm i\left[\theta_\mathrm{V}
+\varepsilon(x)\theta_\mathrm{A}\right]\cdot\tau}\chi(x)$ and 
$\bar\chi(x)\mapsto\bar\chi(x)\mathrm e^{-\mathrm i\left[\theta_\mathrm{V}
-\varepsilon(x)\theta_\mathrm{A}\right]\cdot\tau}$ 
with $\tau\in\mathrm{U}(2)$.

After integrating out the spatial link variable $U_j$ in the leading
order of $1/d$ expansion and introducing the auxiliary fields, 
we have the following partition function:
\begin{align}
 Z&=\int\mathcal{D}[U_0]\,\mathcal{D}[\chi,\bar\chi]\,
 \mathcal{D}[\sigma]\,\mathrm{e}^{-S'[U_0,\chi,\bar\chi;\sigma]}\,,
\end{align}
with
\begin{align}
 \begin{split}
  S'[U_0,\chi,\bar\chi;\sigma]
  =&\sum_{\alpha=\mathrm{u},\mathrm{d}}
  \sum_{x,y}\bar\chi_\alpha(x)\left[m\delta_{x,y}
  +\frac{1}{2}\left\{\mathrm{e}^{\mu_\alpha}U_0(x)\delta_{y,x+\hat0}
  -\mathrm{e}^{-\mu_\alpha}U_0^\dagger(x)
  \delta_{y,x-\hat0}\right\}\right]\chi_\alpha(y)\\  
  &+\sum_{\alpha,\beta=\mathrm{u},\mathrm{d}}\sum_{x,y}
  \left[\sigma_{\beta\alpha}(x)V_\mathrm{M}(x,y)  
  \sigma_{\alpha\beta}(y)+2\sigma_{\beta\alpha}(x)V_\mathrm{M}(x,y)
  M_{\alpha\beta}(y)\right]\,. 
 \end{split}
 \label{eq:prime_action-iso}
\end{align}
Here the subscripts $\alpha, \beta$ represent the species of staggered
fermion, taking ``u'' (for up quarks) and ``d'' (for down quarks). 
$\mu_\mathrm{u}$, $\mu_\mathrm{d}$ are the quark chemical potentials for 
each species of staggered fermion.
$\sigma_{\alpha\beta}(x)$ are auxiliary fields
for the mesonic composites
$M_{\alpha\beta}(x)=\bar\chi_\alpha^a(x)\chi_\beta^a(x)/\Nc$, 
and these vacuum expectation values read
$\left\langle\sigma_{\alpha\beta}(x)\right\rangle
 =-\left\langle M_{\alpha\beta}(x)\right\rangle$. 

Now we replace the auxiliary fields $\sigma_\mathrm{uu}(x)$ and
$\sigma_\mathrm{dd}(x)$ by spatially uniform condensates of the scalar
mode as $\sigma_\mathrm{uu}(x)=\sigma_\mathrm{u}$, 
$\sigma_\mathrm{dd}(x)=\sigma_\mathrm{d}$. 
Then we find that $\sigma_\mathrm{u}$ and $\sigma_\mathrm{d}$
correspond to the chiral condensates of the up and down quarks:
$\sigma_\mathrm{u}
=-\left\langle\bar{\mathrm{u}}^a(x)\mathrm{u}^a(x)\right\rangle/\Nc$ and  
$\sigma_\mathrm{d}
=-\left\langle\bar{\mathrm{d}}^a(x)\mathrm{d}^a(x)\right\rangle/\Nc$. 
We also replace $\sigma_\mathrm{ud}(x)$ and
$\sigma_\mathrm{du}(x)$ by spatially uniform condensates of the
pseudoscalar mode as 
$\sigma_\mathrm{ud}(x)=\mathrm{i}\varepsilon(x)\pi$, 
$\sigma_\mathrm{du}(x)=\mathrm{i}\varepsilon(x)\pi^*$.
Then $\pi$ corresponds to the charged pion condensates: 
$\pi=\left\langle\mathrm{i}\,\bar{\mathrm{u}}^a(x)
\varepsilon(x)\mathrm{d}^a(x)\right\rangle/\Nc$ and 
$\pi^*=\left\langle\mathrm{i}\,\bar{\mathrm{d}}^a(x)
\varepsilon(x)\mathrm{u}^a(x)\right\rangle/\Nc$.

Substituting these mean field values and Eqs.~(\ref{eq:gauge_choice}), 
(\ref{eq:fourier_trans}) in the action
(\ref{eq:prime_action-iso}), we obtain
\begin{align}
 S'[\phi,\chi,\bar\chi;\sigma_\mathrm{u},\sigma_\mathrm{d},\pi]
 =\sum_x\frac{\Nc d}{4}
 \left(\sigma_\mathrm{u}^{\,2}+\sigma_\mathrm{d}^{\,2}
 +2\left|\pi\right|^2\right) 
 -\sum_{\vec{x}}\sum_{a}\sum_{m,n}\bar{X}(m,\vec{x})
 \,G^{-1}(m,n;\vec{x},\phi_a)\,X(n,\vec{x})\,,
\end{align}
where
\begin{align}
 \bar{X}(m,\vec{x})=\left(\tilde{\bar{\mathrm{u}}}(m,\vec{x}),
 \tilde{\bar{\mathrm{d}}}(m,\vec{x})\right)\qquad\text{and}\qquad
 X(n,\vec{x})=
 \begin{pmatrix}
  \tilde{\mathrm{u}}(n,\vec{x})\\\tilde{\mathrm{d}}(n,\vec{x})
 \end{pmatrix}\,,\qquad
\end{align}
and
\begin{align}
 G^{-1}(m,n;\vec{x},\phi_a)=
 \begin{pmatrix}
  -\left[M_\mathrm{u}+\mathrm{i}\sin\left(k_m+\frac{\phi_a(\vec{x})}
  {N_\tau}-\mathrm{i}\mu_\mathrm{u}\right)\right]\delta_{mn}
  &\mathrm{i}\varepsilon(\vec{x})\frac d2\pi^*\delta_{n,m-N_\tau/2}\\
  \mathrm{i}\varepsilon(\vec{x})\frac d2\pi\delta_{n,m-N_\tau/2}
  &-\left[M_\mathrm{d}+\mathrm{i}\sin\left(k_m+\frac{\phi_a(\vec{x})}
  {N_\tau}-\mathrm{i}\mu_\mathrm{d}\right)\right]\delta_{mn}
 \end{pmatrix}\,.
\end{align}
$M_\mathrm{u}=m+(d/2)\sigma_\mathrm{u}$ and 
$M_\mathrm{d}=m+(d/2)\sigma_\mathrm{d}$ are the dynamical quark masses
of up and down quarks. After performing the integration over the
Grassmann variables $X$ and $\bar{X}$, we obtain the following effective
free energy:
\begin{align}
 F_{\mathrm{eff}}
 [\sigma_\mathrm{u},\sigma_\mathrm{d},\pi;T,\muB,\muI]
 =\frac{\Nc d}{4}\left(\sigma_\mathrm{u}^{\,2}
 +\sigma_\mathrm{d}^{\,2}+2\left|\pi\right|^2\right)
 -T\log\left\{\int\mathrm{d}U_0\,\prod_{a=1}^{\Nc}
 \mathrm{Det}\left[G^{-1}(\phi_a)\right]\right\}\,,
 \label{eq:free_energy-iso}
\end{align}
where
\begin{align}
 \mathrm{Det}\left[G^{-1}(m,n;\vec{x},\phi_a)\right]
 =\prod_{n=1}^{N_\tau}\left[\left(\frac d2\right)^2\left|\pi\right|^2
 +\left\{M_\mathrm{u}+\mathrm{i}\sin
 \left(\bar k_n-\mathrm{i}\mu_\mathrm{I}/2\right)\right\}\cdot
 \left\{M_\mathrm{d}-\mathrm{i}\sin
 \left(\bar k_n+\mathrm{i}\mu_\mathrm{I}/2\right)\right\}\right]
\label{eq:determinant-iso}
\end{align}
with $\bar k_{n}=k_n+\phi_a(\vec{x})/N_\tau-\mathrm{i}\muB/\Nc$:
Here we have rewritten the chemical
potentials of the up and down quarks as $\muB$ and $\mu_\mathrm{I}$ using
the definition $\mu_\mathrm{u}=\muB/\Nc+\mu_\mathrm{I}/2$ and
$\mu_\mathrm{d}=\muB/\Nc-\mu_\mathrm{I}/2$. 
The integration over $U_0$ in the Polyakov gauge is defined in
Appendix~\ref{app:integration}.  

It is impossible to perform the summation over $n$ in
Eq.~(\ref{eq:determinant-iso}) analytically for general $\muB$ and
$\muI$. However, we can do it in two interesting cases. One is
the case that $\muI$ is lower than a critical
value $\muI^\mathrm{cri}\sim m_\pi$, where the pion does not
condense, $\pi=0$ \cite{son01}. This case is  
relevant to realistic systems such as heavy ion collisions and
electroneutral neutron stars. The other is the case that $\muB=0$, 
which means we can put $\sigma_\mathrm{u}=\sigma_\mathrm{d}$. 
In this case, we can compare our results with those obtained by the
recent Monte Carlo lattice simulations \cite{kogut02}.

\subsection{The phase structure for $\mu_\mathrm{I}<m_\pi$ 
  \label{sec:isospin_case1}} 

First, we derive the analytical expression for the effective free energy 
in the case of $\mu_\mathrm{I}<\mu_\mathrm{I}^\mathrm{cri}$, where we
can put $\pi=0$. The proof for the existence of such a critical value
and its correspondence to the pion mass in the framework of
strong coupling lattice QCD will be given in the next subsection.
In this case, we can calculate the product over $n$ using
the formula in Appendix~\ref{app:summation}:
\begin{align}
\begin{split}
 \mathrm{Det}\left[G^{-1}(m,n;\vec{x},\phi_a)\right]
 &=\prod_{n=1}^{N_\tau}\left\{M_\mathrm{u}+\mathrm{i}\sin
 \left[k_n+\phi_a(\vec{x})/\Nt-\mathrm{i}\mu_\mathrm{u}\right]\right\}\cdot
 \left\{M_\mathrm{d}+\mathrm{i}\sin
 \left[k_n+\phi_a(\vec{x})/\Nt-\mathrm{i}\mu_\mathrm{d}\right]\right\}\\
 &=\left\{2\cosh\left[N_\tau E_\mathrm{u}\right]+2\cos[\phi_a(\vec{x})
 -\mathrm{i}N_\tau\mu_\mathrm{u}]\right\}\cdot
 \left\{2\cosh\left[N_\tau E_\mathrm{d}\right]+2\cos[\phi_a(\vec{x})
 -\mathrm{i}N_\tau\mu_\mathrm{d}]\right\}
\end{split}
\label{eq:up-down_sectors}
\end{align}
with one-dimensional up and down quark excitation energy 
$E_\mathrm{u}=\arcsinh M_\mathrm{u}$, 
$E_\mathrm{d}=\arcsinh M_\mathrm{d}$.

Using the formula in Appendix~\ref{app:integration}, we can complete the 
$\SU(\Nc)$ integration over $U_0$ in
Eq.~(\ref{eq:free_energy-iso}) for general $\Nc$. Then we obtain the
analytical expression for the effective free energy as follows:
\begin{align}
 F_{\mathrm{eff}}[\sigma_\mathrm{u},\sigma_\mathrm{d};
 T,\muB,\muI<\mu_\mathrm{I}^\mathrm{cri}]
 =\frac{\Nc d}{4}
 \left(\sigma_\mathrm{u}^{\,2}+\sigma_\mathrm{d}^{\,2}\right)
 -T\log\left\{\sum_{n}\det_{i,j}Q_{n+i-j}\right\}\,,
 \label{eq:free_energy1-iso}
\end{align}
where
\begin{align}
\begin{split}
 Q_0&=4\cosh\left[E_\mathrm{u}/T\right]\cdot\cosh\left[E_\mathrm{d}/T\right] 
 +2\cosh\left[(\mu_\mathrm{u}-\mu_\mathrm{d})/T\right]\,,\\
 Q_{\pm1}&=2\cosh\left[E_\mathrm{u}/T\right]\cdot
 \left(\cosh\left[\mu_\mathrm{d}/T\right]
 \pm\sinh\left[\mu_\mathrm{d}/T\right]\right)
 +2\cosh\left[E_\mathrm{d}/T\right]\cdot
 \left(\cosh\left[\mu_\mathrm{u}/T\right]
 \pm\sinh\left[\mu_\mathrm{u}/T\right]\right)\,,\\ 
 Q_{\pm2}&=\cosh\left[(\mu_\mathrm{u}+\mu_\mathrm{d})/T\right]
 \pm\sinh\left[(\mu_\mathrm{u}+\mu_\mathrm{d})/T\right]\,,\qquad
 Q_{|n|\geq3}=0\,,
\end{split}
\end{align}
and the determinant is to be taken with respect to
$i,j=1,2,\dots,\Nc$. 

Although the effective free energy has a complicated expression,
it reduces to a simple form at $T=0$. In the zero temperature limit
$\Nt\to\infty$, the contribution of $\phi_a(\vec{x})$ in
Eq.~(\ref{eq:up-down_sectors}) vanishes. Then the up quarks and the
down quarks are completely decoupled, and we can write the
effective free energy for $T=0$ as 
\begin{align}
 F_{\mathrm{eff}}[\sigma;T=0,\muB,\muI<\mu_\mathrm{I}^\mathrm{cri}]
 =\left[\frac{\Nc d}{4}\sigma_\mathrm{u}^{\,2}
 -\max\left\{\Nc\mu_\mathrm{u},\Nc E_\mathrm{u}\right\}\right]
 +\left[\frac{\Nc d}{4}\sigma_\mathrm{d}^{\,2}
 -\max\left\{\Nc\mu_\mathrm{d},\Nc E_\mathrm{d}\right\}\right]\,.
\end{align}
Each term coincides with the effective free energy at $T=0$ for $\Nf=4$  
[Eq.~(\ref{eq:T=0})] with quark chemical potential $\mu_\mathrm{u}$ or 
$\mu_\mathrm{d}$. Therefore the discussions on chiral restoration at
finite density given in Sec.~\ref{sec:baryon_analytical} hold true in
each sector.

\subsubsection*{The phase diagrams}

\begin{figure}[tp]
 \begin{center}
  \includegraphics[width=0.98\textwidth,clip]{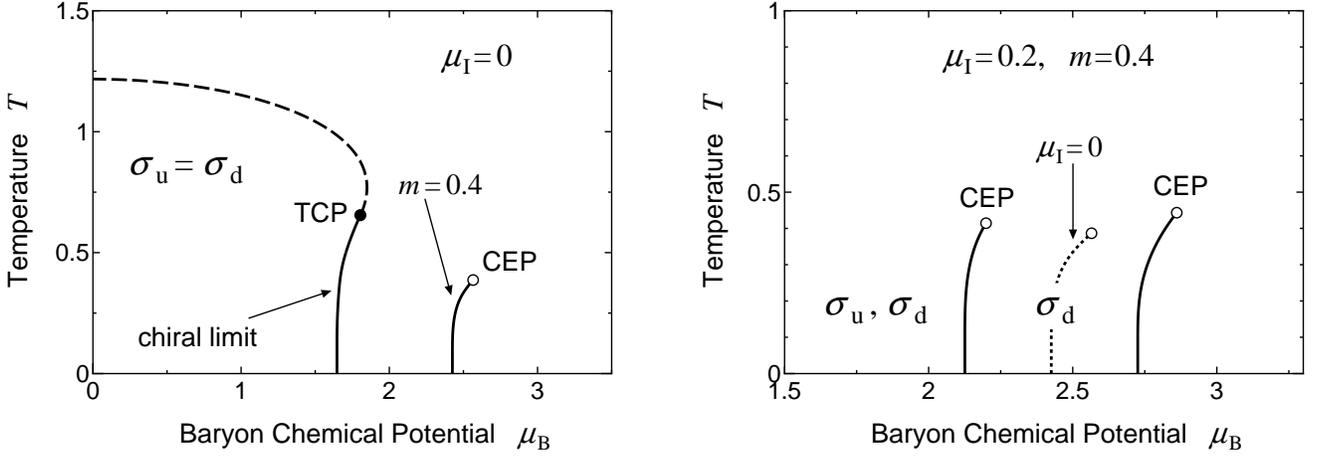}
  \caption{Phase diagrams of strong coupling lattice QCD with
  $\Nc=3,\Nf=8$ in the $T$-$\muB$ plane with zero (the left panel) and
  finite (the right 
  one) isospin chemical potential $\mu_\mathrm{I}$. In the left panel, 
  the solid (dashed) line denotes the phase transition line for the first
  (second) order phase transition in the chiral limit at $\muI=0$. The
  first order line and its critical end point for $m=0.4$ are also shown. 
  The right panel shows the phase diagram at $\muI=0.2$ and $m=0.4$. The
  two solid lines represent the phase transition lines for the first
  order phase transitions associated with up (the left line) and down
  (the right one) quark condensates, respectively. The first order phase 
  transition line for $\muI=0$ is plotted by the dotted line for
  comparison. The symbol $\sigma_\mathrm{u}$ or $\sigma_\mathrm{d}$ in
  the figure represents the region where the condensate has a large
  value. \label{fig:PD_baryon}}
 \end{center}
\end{figure}

At finite temperature, the integration over the temporal gauge link 
variable $\phi_a$ of Eq.~(\ref{eq:up-down_sectors}) nontrivially
couples the up and down quark sectors. Let us see its consequences
numerically here. We show in Fig.~\ref{fig:PD_baryon} the phase diagram
in the plane of $T$ and $\muB$ with $\Nf=8$, determined by minimizing 
the effective free energy Eq.~(\ref{eq:free_energy1-iso}) for $\Nc=3$,
$d=3$ with respect to the order parameters $\sigma_\mathrm{u}$,
$\sigma_\mathrm{d}$. 

First we consider the case of $\muI=0$ in which we can put
$\sigma_\mathrm{u}=\sigma_\mathrm{d}$; the phase 
diagram is shown in the left panel of Fig.~\ref{fig:PD_baryon}. 
This should be compared to the case of four-flavor QCD in
Fig.~\ref{fig:phase_diagram}. 
The outline of the phase diagrams is quite consistent, and the discussions
given in Sec.~\ref{sec:baryon_numerical} hold true here, while the
critical temperature for $\Nf=8$ flavor is lower than that for $\Nf=4$. 
This is expected from a physical point of view: 
With more flavors, the chiral condensate is broken faster by their
thermal excitations. In other words, the critical temperature decreases
on increasing the number of quark flavors. 
What we want to emphasize here is that the integration over $\phi_a$ 
causes the $N_\mathrm{f}$ dependence of the critical temperature even at 
the mean field level.

Next we consider the phase diagram at $\muI=0.2$ with $m=0.4$. 
As shown in the right panel 
of Fig.~\ref{fig:PD_baryon}, the introduction of $\muI$ splits the first
order phase transition line into two lines 
associated with the jump of the chiral condensate of the up quark and
that of the down quark. They shift in opposite directions of $\muB$
and terminate at the second order critical end points. Such a successive
chiral restoration with finite $\muI$ on increasing
$\muB$ is easy to understand intuitively 
from the fact that up and down quarks experience the chemical potential 
differently, $\mu_\mathrm{u}=\muB/\Nc+\muI/2$ and 
$\mu_\mathrm{d}=\muB/\Nc-\muI/2$.

What is nontrivial in our result is the position of the critical
end points. The temperatures of the critical end points for
$\sigma_\mathrm{u}$ and $\sigma_\mathrm{d}$ are both higher than that in
the case of $\muI=0$. 
This is in contrast to studies using the random matrix model
\cite{klein03}, Nambu$-$Jona-Lasinio model \cite{frank03} and ladder QCD 
approach \cite{barducci03}, where the temperature of the critical end
points is not affected by $\muI$ because the up quark sector and the
down quark sector are completely uncoupled in such approaches. 
In our approach, as discussed above, the up and down 
quarks are nontrivially coupled via the integration of the temporal
gauge link variable $U_0$ at finite temperature.

\subsection{The phase structure for $\muB=0$
  \label{sec:isospin_case2}}

We derive the analytical expression for the effective free
energy in the case of $\muB=0$. In this case, we can put
$\sigma=\sigma_\mathrm{u}=\sigma_\mathrm{d}$ and obtain 
the following expression for Eq.~(\ref{eq:determinant-iso}):
\begin{align}
\begin{split}
 \mathrm{Det}\left[G^{-1}(m,n;\vec{x},\phi_a)\right]
 &=\prod_{n=1}^{N_\tau}\left[\left(\frac d2\right)^2\left|\pi\right|^2
 +\left\{M+\mathrm{i}\sin\left(k_n+\frac{\phi_a(\vec{x})}
 {N_\tau}-\mathrm{i}\frac{\muI}2\right)\right\}\cdot
 \left\{M-\mathrm{i}\sin\left(k_n+\frac{\phi_a(\vec{x})}
 {N_\tau}+\mathrm{i}\frac{\muI}2\right)\right\}\right]\,.
\end{split}
\end{align}
This expression has the same form as the determinant calculated in
\cite{nishida03} for $\SUc(2)$ QCD. Therefore we can use the formula in
the Appendix of \cite{nishida03} for the product over the Matsubara
frequencies, and then we obtain 
\begin{align}
 \mathrm{Det}\left[G^{-1}(m,n;\vec{x},\phi_a)\right]
 =\left\{2\cosh\left[N_\tau E_-\right]
 +2\cos\left[\phi_a(\vec{x})\right]\right\}\cdot
 \left\{2\cosh\left[N_\tau E_+\right]
 +2\cos\left[\phi_a(\vec{x})\right]\right\}
 \label{eq:quark-antiquark_sectors}
\end{align}
with one-dimensional up quark and down antiquark 
(or up antiquark and down quark) excitation energy 
\begin{align}
 E_\pm
 =\mathrm{arccosh}\left(\sqrt{(1+M^2)\cosh^2\left[\mu_\mathrm{I}/2\right] 
 +(d/2)^2|\pi|^2}\pm M\sinh\left[\mu_\mathrm{I}/2\right]\right)\,.
\end{align}

Using the formula in Appendix~\ref{app:integration}, we can complete the 
$\mathrm{SU}(\Nc)$ integration over $U_0$ in
Eq.~(\ref{eq:free_energy-iso}) for general $\Nc$. Then we obtain the
analytical expression for the effective free energy as follows:
\begin{align}
 F_{\mathrm{eff}}[\sigma,\pi;T,\muB=0,\mu_\mathrm{I}]
 =\frac{\Nc d}{2}\left(\sigma^2+\left|\pi\right|^2\right)
 -T\log\left\{\sum_{n}\det_{i,j}R_{n+i-j}\right\}\,,
  \label{eq:free_energy2-iso}
\end{align}
where
\begin{align}
\begin{split}
 R_0&=4\cosh\left[E_-/T\right]\cdot\cosh\left[E_+/T\right]+2\,,\\
 R_{\pm1}&=2\cosh\left[E_-/T\right]+2\cosh\left[E_+/T\right]\,,\\
 R_{\pm2}&=1\,,\qquad R_{|n|\geq3}=0\,,
\end{split}
\end{align}
and the determinant is to be taken with respect to
$i,j=1,2,\dots,\Nc$. Note that for $m=0$ and $\muI=0$, the effective
free energy is a function of only $\sigma^2+|\pi|^2$. Of course, 
this is from the chiral symmetry $\mathrm{U}(2)_\mathrm{A}$ of the
original action at $m=\muI=0$. The symmetry between $\sigma$ and $\pi$
will play an important role in understanding numerical results on the
phase structure.

Although the effective free energy has a complicated expression,
it reduces to a simple form at $T=0$. In the zero temperature limit
$\Nt\to\infty$, the contribution of $\phi_a(\vec{x})$ in
Eq.~(\ref{eq:quark-antiquark_sectors}) vanishes, and we obtain
the effective free energy for $T=0$ as
\begin{align}
 F_{\mathrm{eff}}[\sigma,\pi;T=0,\muB=0,\mu_\mathrm{I}]
 =\Nc\left[\frac{d}{2}
 \left(\sigma^2+\left|\pi\right|^2\right)
 -\left(E_-+E_+\right)\right]\,.
 \label{eq:free_energy2-isoT0}
\end{align}
If we replace $\pi$ by $\Delta$ and $\muI$ by $\muB$, the above 
expression corresponds to the free energy at $T=0$ of $\SUc(2)$ QCD
studied in \cite{nishida03} up to the overall factor $\Nc$. This
correspondence is rather natural from the point of view of the symmetry
and the effect of the chemical potential. In the present system, up and
down quarks are indistinguishable because of chiral symmetry at $\muI=0$
and they undergo the effect of $\muI$ with opposite signs. On the other
hand, in $\SUc(2)$ QCD, the quark and antiquark are indistinguishable
because of the Pauli-G\"{u}rsey symmetry \cite{pauli57} at $\muB=0$, and 
they also undergo the effect of $\muB$ with opposite signs. Therefore
there is a correspondence between the up (down) quark in the finite
isospin density system and the (anti)quark in the finite baryon density
$\SUc(2)$ system, which also means a correspondence between the pion
condensation and the diquark condensation. Note that the difference of
the effective free energy in the two systems at $T\neq0$ is due to the
integration of $U_0$ for the finite isospin density system, which
couples up and down quarks nontrivially at finite temperature. 

As calculated in \cite{nishida03} for the case of $\SUc(2)$ QCD, we
can derive analytical properties from the effective free energy at 
$T=0$ in Eq.~(\ref{eq:free_energy2-isoT0}). The system has two
critical chemical potentials, the lower one
$\mu_\mathrm{c}^\mathrm{low}$ and the upper one
$\mu_\mathrm{c}^\mathrm{up}$, given, respectively, by 
\begin{align}
 \mu_{\mathrm{c}}^{\mathrm{low}}=2\,\mathrm{arccosh}\sqrt{1+mM}
 \quad\text{and}\quad
 \mu_{\mathrm{c}}^{\mathrm{up}}=2\,\mathrm{arccosh}\sqrt{1+K^2}\,,
\label{eq:threshold}
\end{align}
where $M$ is a solution of the chiral gap equation with quark mass $m$,
and we have defined $K$ as the solution of the equation
\begin{align}
 \frac{2}{d}\left(K-\frac{m^2}{K}\right)=(1+K^2)^{-1/2}\,.
\end{align}
The nonvanishing value of the pion condensate $\pi\neq0$ is possible only
for
$\mu_{\mathrm{c}}^{\mathrm{low}}\le\muI\le\mu_{\mathrm{c}}^{\mathrm{up}}$.  
As long as $\muI<\mu_{\mathrm{c}}^{\mathrm{low}}$, 
the empty vacuum gives $\rhoI=-\partial F_\mathrm{eff}/\partial\muI=0$
and $\pi=0$. On the other hand, 
for $\muI>\mu_{\mathrm{c}}^{\mathrm{up}}$, saturation of the isospin 
density occurs, leading to $\rhoI=\Nc$ and $\pi=0$.
For sufficiently small $m$, the lower critical chemical potential
reduces to 
\begin{align}
 \mu_{\mathrm{c}}^{\mathrm{low}} 
 =2\,m^{1/2}\cdot\left\{\frac{(1+d^2)^{1/2}-1}{2}\right\}^{1/4}\,.
\end{align}
This expression can be rewritten as
$\mu_{\mathrm{c}}^{\mathrm{low}}=m_{\pi}$ with $m_{\pi}$ being the pion
mass obtained from the excitation spectrum in the vacuum
\cite{kluberg-stern83} up to leading order of the $1/d$
expansion. This is nothing but a proof of the fact that the critical
value of the isospin chemical potential $\mu_\mathrm{I}^\mathrm{cri}$
assumed in the previous subsection indeed exists and corresponds to the
pion mass. 

\subsubsection*{Condensates and isospin density}

\begin{figure}[tp]
 \begin{center}
  \includegraphics[width=0.95\textwidth,clip]{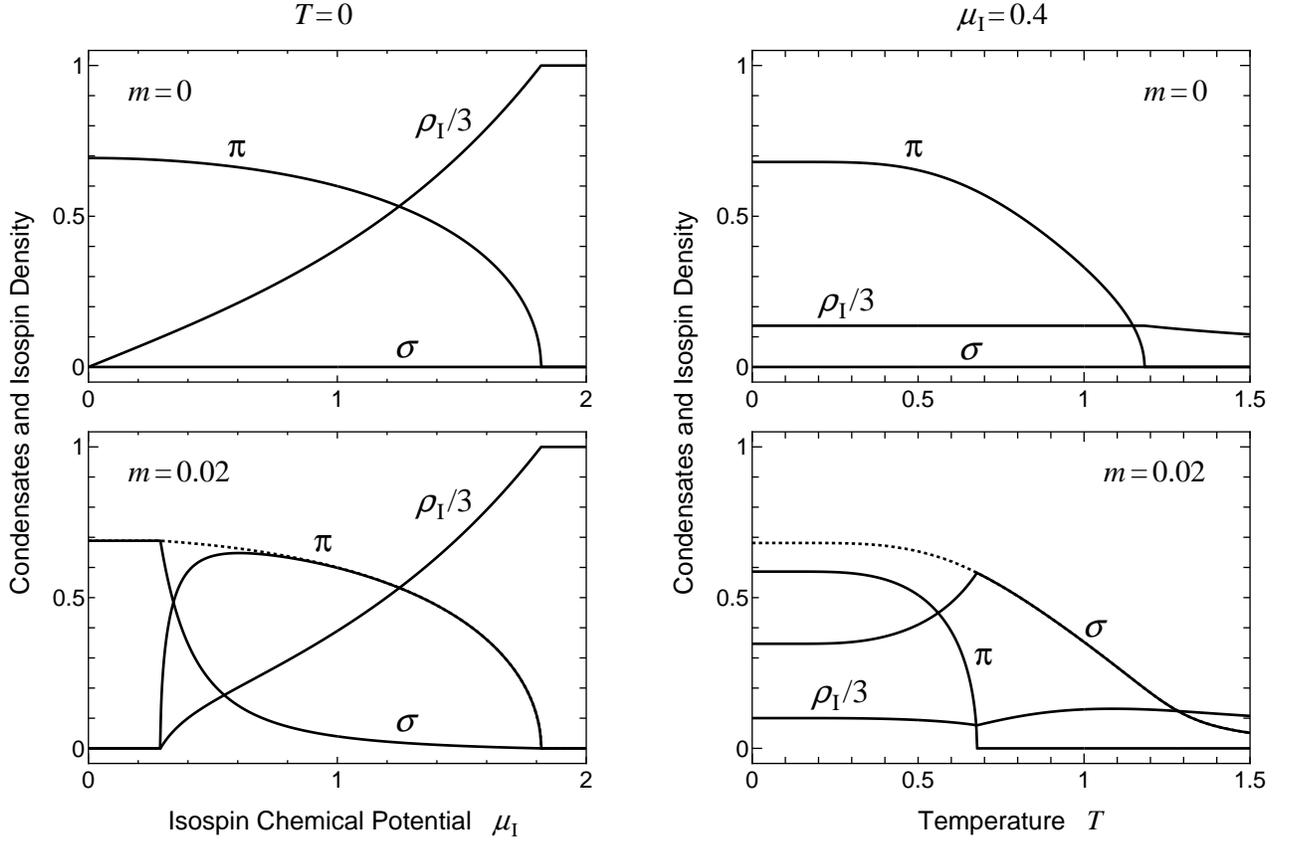}
  \caption{Chiral condensate $\sigma$, pion condensate $\pi$ and isospin
  density $\rho_\mathrm{I}$ for $\Nc=3,\Nf=8$. In the left panels, they
  are plotted as functions of the isospin chemical potential $\muI$ at
  $T=0$ for $m=0$ (the upper panel) and $m=0.02$ (the lower one). In the
  right panels, they are plotted as functions of the temperature $T$ at
  $\muI=0.4$ for $m=0$ (the upper panel) and $m=0.02$ (the lower
  one). The dotted lines in the lower panels indicate the total magnitude
  of the condensates $\sqrt{\sigma^2+\pi^2}$. 
  \label{fig:condensates}}
 \end{center}
\end{figure}

Now we determine the chiral condensate $\sigma$ and the pion condensate
$\pi$ numerically by minimizing the effective free energy with $\Nf=8$
flavors in Eq.~(\ref{eq:free_energy2-iso}) for $\Nc=3$, $d=3$. The
isospin density $\rho_\mathrm{I}=-\partial F_\mathrm{eff}/\partial\muI$
is also calculated as a function of $\muI$ and $T$. The results are
shown in Fig.~\ref{fig:condensates} for $m=0$ and $m=0.02$. Here we
choose $\pi$ to be real.

Let us first consider $\sigma$, $\pi$ and $\rhoI$ as functions of $\muI$
at $T=0$, shown in the left panels of Fig.~\ref{fig:condensates}. 
In the chiral limit $m=0$ (the upper left panel), $\sigma$
is always zero, while $\pi$ decreases monotonically as
a function of $\muI$ and shows a second order transition when $\muI$
becomes of order 2. On the other hand, $\rho_\mathrm{I}$ increases 
until the saturation point $\rhoI=3$ where quarks occupy the maximally 
allowed configurations by the Fermi statistics.

The lower left panel of Fig.~\ref{fig:condensates} is the result for
small quark mass $m=0.02$. As we discussed above, 
there exists a lower critical chemical potential
$\mu_{\mathrm{c}}^{\mathrm{low}}\sim m_\pi$ given by
Eq.~(\ref{eq:threshold}). Both $\pi$ and $\rho_\mathrm{I}$ start to take  
finite values only for $\muI>\mu_{\mathrm{c}}^{\mathrm{low}}$. 
One can view the behavior of $\sigma$ and $\pi$ with
finite quark mass as the manifestation of two different mechanisms. One  
is a continuous ``rotation'' from chiral condensation to pion
condensation above $\muI=\mu_{\mathrm{c}}^{\mathrm{low}}$ with 
$\sqrt{\sigma^2+\pi^2}$ varying smoothly. The other is the 
``saturation effect''; the isospin density forces the pion condensate to 
decrease and disappear for large $\muI$ as seen in the previous case of
$m=0$. 

The ``rotation'' can be understood as follows. As we have discussed, 
the effective free energy at vanishing $m$ and $\muI$ has a chiral
symmetry between the chiral condensate and the pion condensate. 
The effect of $m$ ($\muI$) is to break this symmetry in the direction of
the chiral (pion) condensation favored. Therefore at finite $m$  
a relatively large $\sigma$ appears predominantly for the small
$\muI$ region. Once $\muI$ exceeds a threshold value 
$\mu_{\mathrm{c}}^{\mathrm{low}}$, $\sigma$ decreases  
while $\pi$ increases, because the effect of $\muI$
surpasses that of $m$. As $\muI$ becomes larger, the pion condensate
begins to decrease in turn by the effect of the ``saturation''
and eventually disappears when $\muI$ exceeds the upper critical value
$\mu_{\mathrm{c}}^{\mathrm{up}}$ (of the order of 2 for $T=0$). 
Such a behavior of $\pi$ is also observed in recent
Monte Carlo simulations of the same system \cite{kogut02}.

Next we consider the chiral and pion condensates as functions of $T$
at $\muI=0.4$ in the right panels of Fig.~\ref{fig:condensates}. 
In the chiral limit $m=0$ (the upper right panel), 
$\sigma$ is always zero, while $\pi$ decreases
monotonically and shows a second order phase transition at some critical 
temperature $T_{\mathrm{c}}$.

For small quark mass $m=0.02$ (the lower right panel), 
both $\sigma$ and $\pi$ have finite values. 
$\pi$ decreases monotonically as $T$ increases and shows a
second order phase transition at a critical temperature smaller than for
$m=0$. On the other hand, $\sigma$ increases as $\pi$
decreases, so that the total condensate $\sqrt{\sigma^2+\pi^2}$ is a
smoothly varying function of $T$. The understanding based on the
approximate symmetry between the chiral and pion condensates is thus
valid. An interesting observation is that $\sigma$,
although it is a continuous function of $T$, has a cusp shape
associated with the phase transition of the pion condensate.

Finally let us compare the results for $m=0$ and those for $m=0.02$
in Fig.~\ref{fig:condensates}. Looking at the two figures 
at the same $T$ or $\muI$, we find that the pion condensate $\pi$ for
$m=0$ and the total condensate $\sqrt{\sigma^2+\pi^2}$ for $m=0.02$ have 
almost the same behavior. This indicates that, although the current
quark mass suppresses $\pi$, the price to pay is an increase in
$\sigma$, so as to  make the total condensate insensitive to the 
presence of small $m$.

\subsubsection*{The phase diagrams}

\begin{figure}[tp]
 \parbox[b]{.48\textwidth}
 {\begin{center}
   \includegraphics[width=.475\textwidth,clip]{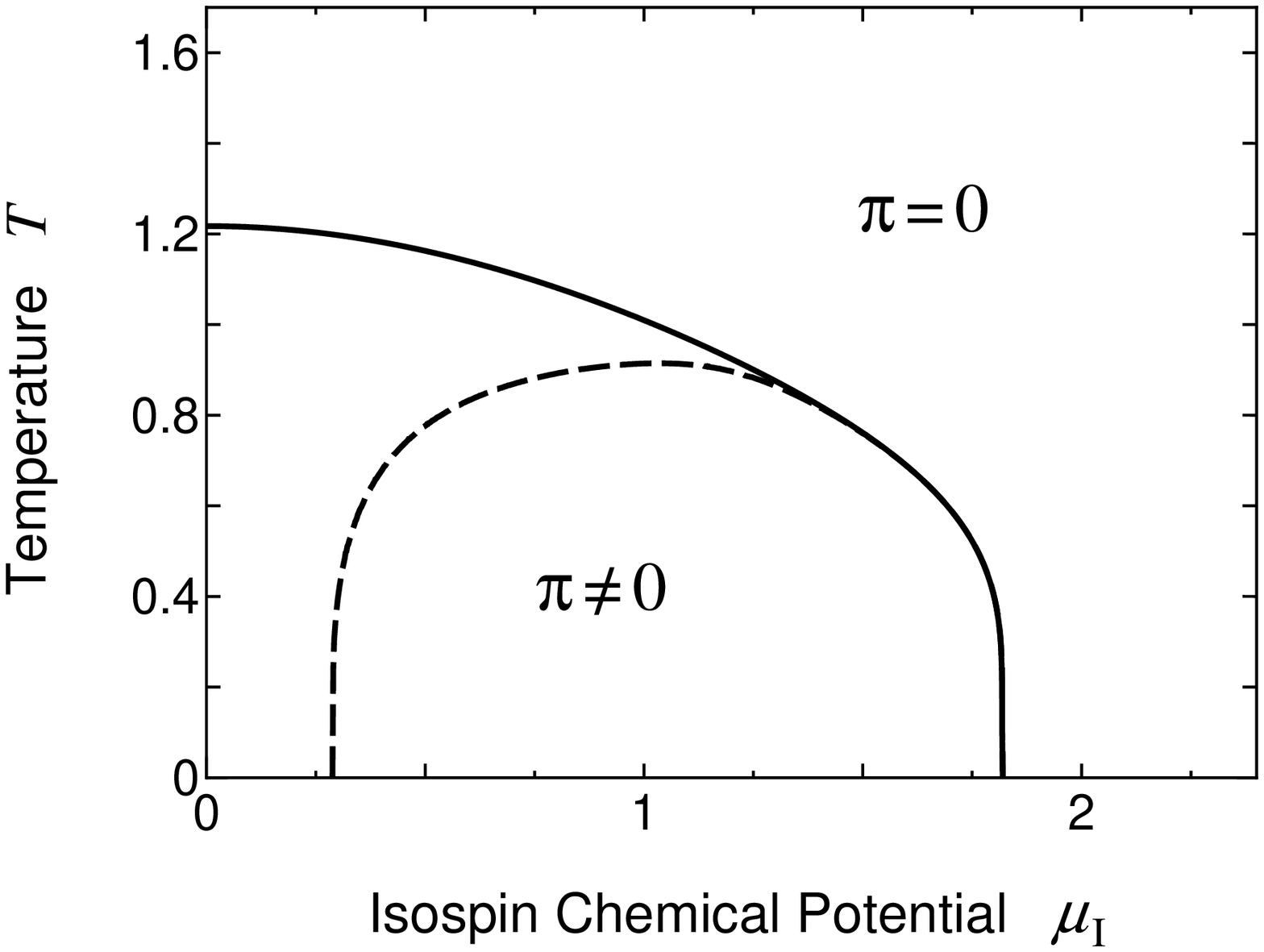}
   \caption{Phase diagram of strong coupling lattice QCD with
   $\Nc=3,\Nf=8$ in the $T$-$\muI$ plane. The solid (dashed) line denotes
   the second order critical line for pion condensation for $m=0$ 
   ($m=0.02$). \label{fig:PD_isospin-2D}}
  \end{center}}
 \hfill
 \parbox[b]{.48\textwidth}
 {\begin{center}
   \includegraphics[width=.45\textwidth,clip]{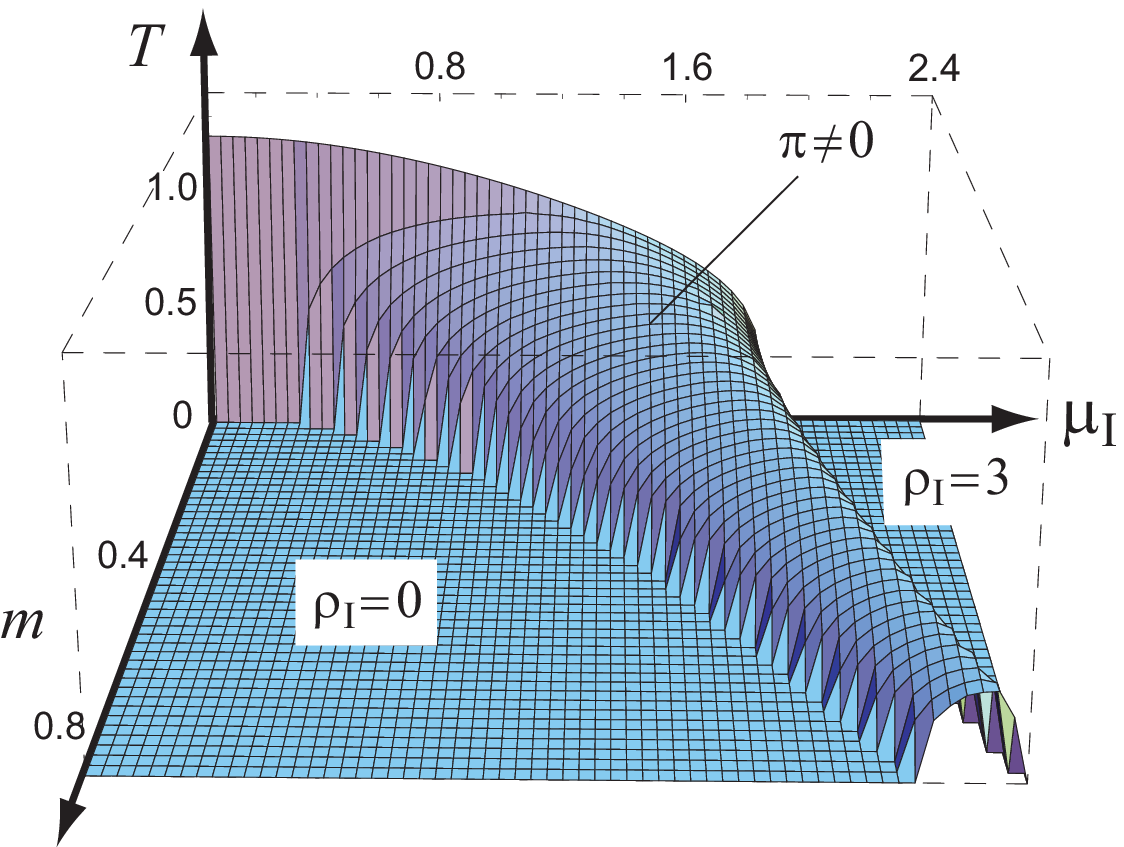}
   \caption{Phase structure of strong coupling lattice QCD with
   $\Nc=3,\Nf=8$ in the $T$-$\muI$-$m$ space. The surface represents the
   critical surface for pion condensation, which separates the
   region where $\pi\neq0$ (inside) from the region where $\pi=0$
   (outside). \label{fig:PD_isospin-3D}}
  \end{center}}
\end{figure}

Now we show in Fig.~\ref{fig:PD_isospin-2D} the phase diagram of the
strong coupling lattice QCD with $\Nc=3,\Nf=8$ in the $T$-$\muI$ plane. 
The solid line denotes the critical line
separating the pion condensation phase $\pi\not=0$ from the normal phase
$\pi=0$ in the chiral limit $m=0$ and the dashed line is for small quark
mass $m=0.02$. The phase transition is of second order on these critical
lines. The chiral condensate $\sigma$ is everywhere zero for $m=0$, 
while it is everywhere finite for $m\neq 0$. 
In the latter case, however, $\sigma$ is particularly large
in the region between the solid line and the dashed line. 

Then the phase structure in the three dimensional
$T$-$\muI$-$m$ space is shown in Fig.~\ref{fig:PD_isospin-3D}. The pion 
condensate has a nonvanishing value inside the critical surface and
the phase transition is of second order everywhere on this critical
surface. The second order phase transition is consistent with other
analyses employing the mean field approximation
\cite{klein03,vanderheyden01}. 
On the other hand, Monte Carlo simulations of QCD with finite $\muI$
\cite{kogut02} show an indication of the presence of a
tricritical point in the $T$-$\muI$ plane at which the property of the
phase transition line changes from second order to first order
as $\muI$ increases. Clarifying what induces such a first order phase
transition is an interesting open problem.

The plane at $T=0$ in Fig.~\ref{fig:PD_isospin-3D} shows the phase
diagram in terms of $\muI$ and $m$. The lower left of the plane
corresponds to the vacuum with no isospin density present, $\rhoI=0$. 
On the other hand, the upper right 
of the plane corresponds to the saturated system, $\rhoI=3$, in which
every lattice site is occupied by three pions. There is a pion
condensation phase (a superfluid phase of isospin charge) with $\pi\not=0$
and $0<\rhoI<3$ bounded by the above two limiting cases. They are
separated by the two critical lines given in Eq.~(\ref{eq:threshold}).

It is worth mentioning here that we can see the corresponding system
in the context of condensed matter physics; the hard-core boson Hubbard
model has a similar phase diagram in which a superfluid phase is
sandwiched by Mott-insulating phases with zero or full density
\cite{schmid01}. Here we can give a physical explanation for the
saturation effect, that is, why the saturated isospin density forces the 
pion condensate to disappear. 
Suppose that one imposes a small external field for the
isospin charge on the system in which every lattice site is occupied
maximally by pions. However, a pion at one site cannot hop to the next
site, owing to the Pauli principle of constituent quarks, and thus the
isospin current never appears. This results in a zero condensate because
the superfluid current is proportional to the square of the absolute
value of the condensate. This is nothing but an analogous phenomenon to
the Mott insulator. 

The fact that $\SUc(3)$ QCD at finite isospin density has
the same phase structure as that of $\SUc(2)$ QCD at finite
baryon density can be understood from the aspect of the symmetry and its 
breaking pattern. $\SUc(3)$ QCD at $\muI=0$ and $m=0$ has chiral
symmetry between the chiral condensate and the pion condensate.  
On the other hand, $\SUc(2)$ QCD at  $\muB=0$ and $m=0$ 
has Pauli-G\"{u}rsey symmetry between the chiral condensate and the
diquark condensate. In both theories, the introduction of $m$ breaks the
symmetries so that the chiral condensate is favored, and the introduction 
of $\muI$/$\muB$ breaks the symmetries so that the pion/diquark
condensate is favored. This is the reason why these two theories have
similar phase structures at finite density.

\section{Summary and discussion}

In this paper, we studied the phase structure of hot and dense QCD
with $\Nc=3$ and $\Nf=4,8$ in the strong coupling limit, formulated on
a lattice. We employed the $1/d$ expansion (only in the spatial
direction) and the mean field approximation to derive the free energy
written in terms of the chiral condensate $\sigma$ and the pion
condensate $\pi$ at finite temperature $T$, baryon chemical potential
$\muB$, isospin chemical potential $\muI$ and quark mass $m$.

In Sec.~\ref{sec:baryon}, we investigated the phase structure with
$\Nf=4$ in the $T$-$\muB$ plane, and the phase transition of chiral
restoration was found to be of second order in the high temperature
region and of first order in the low temperature region. Analytical
formulas for the critical line of the second order transition and the
position of the tricritical point were derived. The critical temperature 
at small quark chemical potential $\mu=\muB/\Nc$ can be expanded as 
$T_\mathrm{c}(\mu)\simeq T_\mathrm{c}(0)-1.5\mu^2/T_\mathrm{c}(0)$,
and the slope was found to be much larger than that calculated in
Monte Carlo lattice simulations
\cite{fodor02,allton02,forcrand02,d'elia03}. We discussed how this
difference could be understood as a result of the strong coupling
limit $g\to\infty$ and the chiral limit $m=0$.  
We also studied the effect of finite quark mass on the phase diagram: 
The position of the critical end point has a relatively large
dependence on the quark mass. In paticular for small $m$, the critical 
end point shifts in the direction of smaller $T$ and smaller $\muB$. 
This is because the gradient of the phase transition line near
the tricritical point is positive; how this feature in the
phase diagram of QCD is affected by the introduction of dynamical
baryons beyond the leading order of the $1/d$ expansion should be
studied in future work.

In Sec.~\ref{sec:isospin}, we derived the free energy including
$\muI$ by extending the formulation studied in Sec.~\ref{sec:baryon}
with two species of staggered fermion ($\Nf=8$).  
First, we observed that the critical temperature of chiral
restoration decreases on increasing the number of quark flavors due to
the thermal excitations of mesons. 
Then we studied the effect of $\muI$ on the phase diagram
in the $T$-$\muB$ plane at finite quark mass: The introduction of $\muI$ 
splits and shifts the first order phase transition line in  opposite
directions of $\muB$. In addition, $\muI$ moves the two critical end
points to higher temperatures than that for $\muI=0$. This is a
nontrivial effect originating from integration over the temporal gauge
link variable at finite temperature, which couples the up and down 
quark sectors. 

Finally, we studied the phase structure in the space of $T$, $\muI$  
and $m$ and found a formal correspondence between color $\SUc(3)$ QCD with
finite isospin density and color $\SUc(2)$ QCD with finite baryon
density. Such a correspondence can be understood by considering the
symmetry and its breaking pattern: chiral (Pauli-G\"{u}rsey) symmetry at 
$m=0$ and $\muI=0$ ($\muB=0$) is broken by the quark mass in the
direction of chiral condensation, while it is broken by the isospin
(baryon) chemical potential in the direction of pion (diquark)
condensation. Although our results are in principle limited to 
strong coupling, the behavior of $\sigma$, $\pi$, and $\rhoI$ in the
$T$-$\muI$-$m$ space has qualitative agreement with recent lattice 
data. We also found that the disappearance of pion condensation 
with saturated density can be understood as a Mott-insulating
phenomenon. 

The remaining work to be explored is to investigate the phase structure
with finite $\muB$ and finite $\muI$ as well as with $T$ and $m$, where
a nontrivial competition between the chiral and pion condensates
takes place. In our formulation, such a study is possible by using the 
effective free energy given in Eq.~(\ref{eq:free_energy-iso}) with
numerical summation over the Matsubara frequency and numerical
integration over the temporal gauge link variable. How the chiral
condensate at finite $\muB$ is replaced by the pion condensate as
$\muI$ increases is one of the most interesting problem to be
examined.

\begin{acknowledgments}
 The author is grateful to T.~Hatsuda, K.~Fukushima and S.~Sasaki for
 continuous and stimulating discussions on this work and 
 related subjects. 
 He also thanks S.~Hands for discussions on the parity
 transformation of staggered fermions. 
\end{acknowledgments}

\appendix

\section{Summation over the Matsubara frequency \label{app:summation}} 

In this appendix, we give the formula for the product over the
Matsubara frequencies of the expression
\begin{align}
 I=\prod_{n=1}^{N_\tau/2}\left[\sin^2\bar k_n+\lambda^2\right]
 =\prod_{n=1}^{N_\tau}\left[\sin^2\bar k_n+\lambda^2\right]^{1/2}
\end{align}
with $\bar k_{n}=2\pi(n-1/2)/\Nt+\phi/N_\tau-\mathrm{i}\mu$ and $\Nt$
an even integer.
Let us first take the logarithm of the above expression 
and differentiate it with respect to $\lambda$:
\begin{align}
 \frac{\partial}{\partial\lambda}\log I
 =\sum_{n=1}^{N_\tau}\frac\lambda{\sin^2\bar k_n+\lambda^2}\,.
\end{align}
Because $\sin(k_n+\phi/N_\tau-\mathrm{i}\mu)$ is invariant
under the shift $n\to n+N_\tau$, we can make the summation of $n$ over
the range $n=-\infty$ to $n=+\infty$ with an appropriate degeneracy
factor $\Omega$. Then the residue theorem enables us to replace the
summation by a complex integral as 
\begin{align}
 \frac{\partial}{\partial\lambda}\log I
 =\frac{1}{\Omega}\left[\oint\frac{\mathrm{d}z}{2\pi\mathrm{i}}
 \frac\lambda{\sin^2(z+\phi/N_\tau-\mathrm{i}\mu)+\lambda^2}
 \frac{-\mathrm{i}N_\tau}{1+\mathrm{e}^{\mathrm{i}N_\tau z}}
 -\sum_{\bar{z}}\frac\lambda{2\sin(\bar z+\phi/N_\tau-\mathrm{i}\mu)
 \cos(\bar z+\phi/N_\tau-\mathrm{i}\mu)}\frac{-\mathrm{i}N_\tau}
 {1+\mathrm{e}^{\mathrm{i}N_\tau\bar{z}}}\right]\,.
 \label{eq:residue_theorem}
\end{align}
Owing to the infinite range of the summation over $n$, we can choose
the closed contour at infinity for the complex integrals with respect
to $z$, and thus such complex integrals go to zero. $\bar{z}$ are the
residues satisfying 
\begin{align}
 \sin^2(\bar z+\phi/N_\tau-\mathrm{i}\mu)+\lambda^2=0\,.
\end{align}
Solving this equation, we obtain
\begin{align}
 \bar z+\phi/N_\tau-\mathrm{i}\mu=\pm\mathrm{i}E+2\pi n\,,\quad
 \pm\mathrm{i}E+\pi+2\pi n\,,
\label{eq:solution}
\end{align}
with $E=\arcsinh\lambda$ and $n=-\infty,\dots,\infty$. Substituting
Eq.~(\ref{eq:solution}) into Eq.~(\ref{eq:residue_theorem}), we obtain 
\begin{align}
\begin{split}
  \frac{\partial}{\partial\lambda}\log I
 &=\frac{1}{\Omega}\sum_{n=-\infty}^\infty\frac1{\cosh E}
 \left[\frac\Nt{1+\mathrm{e}^{-N_\tau E-\mathrm{i}\phi-\Nt\mu}}
 -\frac\Nt{1+\mathrm{e}^{N_\tau E-\mathrm{i}\phi-\Nt\mu}}\right]\\
 &=\frac{\partial E}{\partial\lambda}
 \left[\frac\Nt{1+\mathrm{e}^{-N_\tau E-\mathrm{i}\phi-\Nt\mu}}
 -\frac\Nt{1+\mathrm{e}^{N_\tau E-\mathrm{i}\phi-\Nt\mu}}\right]\\
 &=\frac{\partial}{\partial\lambda}
 \left[\log\left\{\mathrm{e}^{N_\tau E}
 +\mathrm{e}^{-\mathrm{i}\phi-\Nt\mu}\right\}
 +\log\left\{\mathrm{e}^{-N_\tau E}
 +\mathrm{e}^{-\mathrm{i}\phi-\Nt\mu}\right\}\right]\\
 &=\frac{\partial}{\partial\lambda}
 \left(-\mathrm i\phi-\Nt\mu
 +\log\left\{2\cosh\left[N_\tau E\right]
 +2\cos[\phi-\mathrm{i}N_\tau\mu]\right\}\right)\,.
\end{split}
\end{align}
We note that the degeneracy factor $\Omega$ is just canceled by the
infinite degeneracy of the summation on $n$.
After integration with respect to $\lambda$, we find that the
result is expressed in a rather simple form up to irrelevant factors,
\begin{align}
 I=2\cosh\left[N_\tau E\right]+2\cos[\phi-\mathrm{i}N_\tau\mu]\,,
\end{align}
with $E=\arcsinh\lambda$.

\section{$\mathrm{SU}(\Nc)$ integration in the Polyakov gauge
 \label{app:integration}}

In this appendix, we give the formula for the $\mathrm{SU}(\Nc)$ 
integration
\begin{align}
 J=\int\mathrm{d}U_0\,F(U_0)\,.
\end{align}
Here we consider the function that can be written as the product of
$f(\theta_a)$ in the Polyakov gauge as
\begin{align}
 F(U_0)=\prod_{a=1}^{\Nc}f\left(\theta_a\right)\,.
\end{align}
Since the $\mathrm{SU}(\Nc)$ Haar measure in the Polyakov gauge is given
by 
\begin{align}
\begin{split}
  \int\mathrm{d}U_0
 &=\int_{-\pi}^\pi\prod_{a=1}^{\Nc}
 \frac{\mathrm{d}\phi_a}{2\pi}\prod_{a<b}\left.
 \,\left|\mathrm{e}^{\mathrm{i}\phi_a}-
 \mathrm{e}^{\mathrm{i}\phi_b}\right|^2\,\right|_{\sum_a\phi_a=0}\\
 &=\int_{-\pi}^\pi\prod_{a=1}^{\Nc}
 \frac{\mathrm{d}\phi_a}{2\pi}
 \,\left|\epsilon_{i_1\cdots i_{\Nc}}
 \mathrm{e}^{\mathrm{i}\phi_1\left(\Nc-i_1\right)}\cdots
 \mathrm{e}^{\mathrm{i}\phi_{\Nc}\left(\Nc-i_{\Nc}\right)}\right|^2
 \cdot2\pi\,\delta\left(\sum\phi_a\right)\,,
\end{split}
 \label{eq:measure}
\end{align}
the integration becomes
\begin{align}
 J=\int_{-\pi}^\pi\prod_{a=1}^{\Nc}\frac{\mathrm{d}\phi_a}{2\pi} 
 \,\left|\epsilon_{i_1\cdots i_{\Nc}}
 \mathrm{e}^{\mathrm{i}\phi_1\left(\Nc-i_1\right)}\cdots
 \mathrm{e}^{\mathrm{i}\phi_{\Nc}\left(\Nc-i_{\Nc}\right)}\right|^2
 \cdot2\pi\,\delta\left(\sum_{a=1}^{\Nc}\phi_a\right)
 \times\prod_{a=1}^{\Nc}f\left(\theta_a\right)\,.
\end{align}
Because the delta function can be rewritten as
\begin{align}
 2\pi\,\delta\left(\sum_{a=1}^{\Nc}\phi_a\right)
 =\sum_{n=-\infty}^\infty\exp
 \left[-\mathrm{i}n\sum_{a=1}^{\Nc}\phi_a\right]\,, 
\end{align}
we obtain
\begin{align}
 J=\sum_{n=-\infty}^\infty\epsilon_{i_1\cdots i_{\Nc}}
 \epsilon_{j_1\cdots j_{\Nc}}
 \prod_{a=1}^{\Nc}\int_{-\pi}^\pi\frac{\mathrm{d}\phi_a}{2\pi}
 f(\phi_a)\exp\left[-\mathrm{i}\left(n+i_a-j_a\right)\phi_a\right]
 =\Nc!\sum_{n=-\infty}^\infty\det_{i,j} M_{n+i-j}\,.
\end{align}
Here we have defined $M_n$ as
\begin{align}
 M_n=\int_{-\pi}^\pi\frac{\mathrm{d}\phi}{2\pi}
 f(\phi)\exp\left[-\mathrm{i}n\phi\right]\,,
\end{align}
and the determinant is to be taken with respect to
$i,j=1,2,\dots,\Nc$.

\end{document}